\documentclass[%
 aip,
 pof,
 amsmath,amssymb,
 reprint,%
]{revtex4-1}

\usepackage{graphicx}
\usepackage{dcolumn}
\usepackage{bm}

\usepackage[utf8]{inputenc}
\usepackage[T1]{fontenc}
\usepackage{siunitx}
\usepackage{tikz}
\usepackage{subcaption}

\begin{document}

\title[The effect of the Stokes boundary layer on the dynamics of particle pairs in an oscillatory flow]{The effect of the Stokes boundary layer on the dynamics of particle pairs in an oscillatory flow}
\author{T.J.J.M. van Overveld}
\affiliation{Fluids and Flows group and J.M. Burgers Center for Fluid Dynamics, Department of Applied Physics, Eindhoven University of Technology, P.O. Box 513, 5600 MB Eindhoven, The Netherlands}
\author{W.-P. Breugem}%
\affiliation{Laboratory for Aero and Hydrodynamics, Delft University of Technology, 2628 CD Delft, The Netherlands}%
\author{H.J.H. Clercx}
\affiliation{Fluids and Flows group and J.M. Burgers Center for Fluid Dynamics, Department of Applied Physics, Eindhoven University of Technology, P.O. Box 513, 5600 MB Eindhoven, The Netherlands}
\author{M. Duran-Matute}
\email{m.duran.matute@tue.nl}
\affiliation{Fluids and Flows group and J.M. Burgers Center for Fluid Dynamics, Department of Applied Physics, Eindhoven University of Technology, P.O. Box 513, 5600 MB Eindhoven, The Netherlands}

\date{\today}

\begin{abstract}
The alignment of a pair of spherical particles perpendicular to a horizontally oscillating flow is attributed to a non-zero residual flow, known as steady streaming. This phenomenon is the basis of complex patterns in denser systems, such as particle chains and the initial stages of rolling-grain ripples. Previous studies on such self-organization processes used two distinct systems: an oscillating box filled with viscous fluid and an oscillating channel flow, where the fluid oscillates relative to the bottom boundary. 
In this paper, we show that particle pair dynamics in these two systems are fundamentally different, due to the presence of a Stokes boundary layer above the bottom in the oscillating channel flow.
The results are obtained from direct numerical simulations in which the dynamics of a pair of particles are simulated using an immersed boundary method.
The oscillating box and the oscillating channel flow are only equivalent in a limited region of the parameter space, where both the normalized Stokes boundary layer thickness and the normalized relative particle excursion length are small. Overall, the particle dynamics in the oscillating channel flow, compared to the oscillating box, are governed by an additional dimensionless parameter.
\end{abstract}

\maketitle

\section{\label{sec:introduction}Introduction}
Patterns in granular systems have been a subject of study for decades, with applications in both industrial \citep{Jaeger1996} and maritime settings \citep{Blondeaux1990}. Specifically, when the grains are immersed in a fluid (liquid or gas), a rich variety of pattern-forming behavior emerges \citep{aranson2006,sanchez2004,perron2018}.  
This is the case, for example, when a collection of spherical particles is submerged in a viscous fluid and subjected to horizontal oscillations. The particles, then, form chains that are aligned perpendicularly to the oscillating flow \citep{Wunenburger2002,Klotsa2009}. The driving mechanism of the chain-forming phenomenon is the nonzero residual flow around the particles, or `steady streaming' flow, that remains after averaging over a full oscillation period \citep{Riley1966}. 

\citet{Klotsa2007} described the equilibrium state of the system for the shortest possible chains, i.e. a pair of aligned particles. They performed numerical simulations and experiments with pairs of stainless steel spheres in a vibrating box filled with a viscous liquid. They identified that the mean gap between the particles is only a function of the viscous length scale and the streamwise excursion amplitude of the particle relative to the fluid, both normalized by the particle diameter.
Later, \citet{VanOverveld2022} confirmed this finding using theoretical arguments and detailed numerical simulations that show excellent agreement with the experimental data of \citet{Klotsa2007}. In addition, \citet{VanOverveld2022} found two scaling regimes for the mean gap: a viscous-dominated and an advection-dominated regime. It was further shown that the gap between the particles oscillates at twice the driving frequency, with the amplitude of these gap oscillations showing two different scaling regimes just like the mean gap.

Additionally, \citet{Mazzuoli2016} extended the results of \citet{Klotsa2009} towards larger fluid excursion lengths, lower oscillation frequencies, and lower particle densities. This region of the parameter space is more relevant for sand ripple formation under surface gravity waves \citep{Vittori1990}. Such small-scale sediment patterns are important for modeling large-scale morphological processes. These patterns cannot be neglected, since they alter the flow structure and consequently the sediment transport \citep{Thibodeaux1987,Mazzuoli2019}.
Using direct numerical simulations, \citet{Mazzuoli2016} studied the inception of rolling-grain ripples due to steady streaming flows. They showed that the mechanisms for the formation of particle chains are, at the origin, similar to the early stages of rolling-grain ripples. Once the chains are formed, they may be considered as perturbations in the bed morphology from which rolling-grain ripples can further evolve.
In a subsequent series of studies, the formation and dynamics of full rolling-grain ripples were simulated for a larger number of particles \citep{Mazzuoli2019,Vittori2020}. The equilibrium wavelengths of the developed ripples showed good agreement with results from linear stability analysis \citep{Blondeaux1990} and experiments \citep{Rousseaux2004}. Due to computational limitations, it was not possible to obtain functional dependencies for the ripple characteristics as function of the flow conditions. 
Furthermore, extensive validation on the emergence of the particle chains, i.e. the perturbed state from which the patterns may further evolve, remains difficult. There are but few experimental studies in this regime, such as those performed on short chains \citep{Hwang2008} or individual particles in a `U-tube' \citep{Martin1976}. More often, experimental studies focus on large systems instead, containing millions of particles, that have a closer connection to environmental situations \citep{Moosavi2014,Blondeaux2016}. 

It is important to note that \citet{Mazzuoli2016} not only studied a different region of the parameter space, but also a fundamentally different system than \citet{Klotsa2007}. The original work on particle chains by \citet{Klotsa2007} was done in an oscillating, closed box filled with viscous fluid containing a pair of stainless steel spheres. In this system, which we call `oscillating box' from here on, the fluid and all container walls oscillate in unison, i.e. with equal amplitude and frequency. 
Conversely, the system studied by \citet{Mazzuoli2016} is a closer representation of a bed over which an oscillatory flow is induced by gravity waves at a free surface. The velocity difference between bed and bulk flow induces the formation of a Stokes boundary layer above the bed. Alternatively, an oscillating pressure gradient imposed on a fluid in a channel between two horizontal parallel plates yields a similar flow if the distance between the plates is large with respect to the boundary layer thickness. Hence, we refer to this system as `oscillating channel flow' in the remainder of this study.

At first glance, it might seem that the oscillating box and the oscillating channel flow are equivalent and that one could transform from one to the other by a change of reference frame. However, this is not the case. In the oscillating channel flow, there is the streamwise motion of the particles, the (bulk) fluid motion, and the (non-moving) bottom. These can be described as two relative motions: between the fluid and the boundaries, and between the particles and (bulk) flow. Three dimensionless quantities are required to uniquely describe these motions, commonly chosen as: the normalized Stokes boundary layer thickness, the normalized streamwise excursion length of the fluid with respect to the boundaries, and the particle-fluid density ratio \citep{Mazzuoli2016}. 

Contrarily, in the oscillating box, the (bulk) flow and the boundaries move in unison, such that the only relative motion is between the particles and the fluid. Consequently, the streamwise excursion length and particle-fluid density ratio can be replaced by a single dimensionless quantity: the relative excursion length of the particles with respect to the fluid \citep{Klotsa2007,VanOverveld2022}. In fact, \citet{VanOverveld2022} have explicitly shown that the mean state of the system is independent of the particle-fluid density ratio, provided that particles are frictionless. The variation of this density ratio leads to the same scaling relations for the mean gap as a function of the relative excursion length of the particles with respect to the fluid. 

Due to the additional motion between the fluid and the boundaries, we hypothesize that the oscillating channel flow has one additional degree of freedom and that the particle dynamics are governed by an additional dimensionless parameter, compared to the oscillating box. Moreover, the effect of the differences between the systems on the particle dynamics and flow fields is still unknown. Likewise, it is not yet clear to what extent a direct comparison between the two systems, or in other words, between the work by \citet{Klotsa2009} and \citet{Mazzuoli2016}, is valid.

Such knowledge is relevant to determine whether the self-organization in both systems is governed by the same underlying physical mechanisms. In addition, the contribution of the steady streaming flows to pattern formation in environmental settings remains unexplored. By considering only a single pair of particles, i.e. the building block of larger patterns, the underlying physical mechanisms are compared between both systems for different regions of the parameter space.

The aim of this study is to better understand the particle pair dynamics in an oscillating channel flow, by comparing it to the oscillating box. First, we address the differences between both systems in detail, using theoretical arguments. Then, we present results for both systems obtained from direct numerical simulations, in which a pair of particles is simulated using the immersed boundary method (IBM) by \citet{Breugem2012}. The streamwise particle motion is described using its relative excursion length as a function of the flow conditions. The relevant question here is whether the Stokes boundary layer over the bottom significantly affects the streamwise particle motion. Next, we focus on the gap between the particles as a function of the dimensionless parameters governing the problem. The parameter space is explored, including the regions covered by the aforementioned studies of interest \citep{Klotsa2007,Mazzuoli2016,VanOverveld2022}. In particular, we aim at determining if the equilibrium state of the particle pairs exhibits the same two (viscous- and advection-dominated) regimes in both systems. In other words, we want to determine if the Stokes boundary layer over the bottom affects the steady streaming flow and thereby the spanwise particle dynamics. 

\section{\label{sec:formulation}Formulation of the problem and numerical approach}
Both systems consist of an incompressible Newtonian fluid, with density $\rho_f$ and kinematic viscosity $\nu$, between two infinitely large, parallel horizontal plates which are separated by a distance $H'$. Two identical solid spheres with diameter $D$ and density $\rho_s$, such that $\rho_s>\rho_f$, are submerged in the fluid. We assume that the spheres stay in contact with the bottom plate due to gravity, with gravitational acceleration $g$. The Coulomb friction coefficient between the particles and bottom is $\mu_c$. 
We have chosen a right-handed Cartesian coordinate system $(x',y',z')$ with the $y$-axis parallel to the oscillation (streamwise) direction, the $x$-axis in the other horizontal (spanwise) direction, and the $z$-axis pointing upwards, perpendicular to the plates.
The additional relevant variables and parameters are the time $t'$, the local flow velocity $\boldsymbol{u}'=(u',v',w')$, the pressure $p'$, the angular frequency of the oscillating flow $\omega$, the excursion length of the bulk fluid $A'$, and the viscous length scale $\delta'=\sqrt{2\nu/\omega}$.

The variables (and gradient operator $\boldsymbol{\nabla}'$) are made dimensionless using $D$ as typical length scale, $2\pi/\omega$ as typical time scale, and $A'\omega$ as typical velocity scale, as follows:
\begin{eqnarray}\label{eq:nondim}
    &&\left(x,y,z\right)=\frac{\left(x',y',z'\right)}{D}, \quad H=\frac{H'}{D}, \quad A=\frac{A'}{D}, \quad \delta=\frac{\delta'}{D}, \nonumber\\
    &&\boldsymbol{\nabla} = D\boldsymbol{\nabla}', \quad t=\frac{\omega t'}{2\pi}, \quad \boldsymbol{u}=\frac{\boldsymbol{u}'}{A'\omega}, \quad p = \frac{p'}{\rho_f A' D\omega^2}.
\end{eqnarray}
Alternatively, we could have chosen $\delta$ as reference length scale, since variations in the flow fields are typically expected on the scale of the oscillatory boundary layer thickness. This approach is used by e.g. \citet{Mazzuoli2016}. Instead, we follow the more classical approach that was used to fundamentally describe steady streaming flows by \citet{Riley1966}.

\subsection{Fluid motion}
The fluid is driven by an external, oscillating pressure gradient
\begin{equation}\label{eq:externalpressure}
    -\boldsymbol{\nabla}p_e = \cos(2\pi t) \boldsymbol{\hat{y}},
\end{equation}
such that the velocity of the bulk flow, far away from boundaries, is
\begin{equation}
    \boldsymbol{u}_b = \sin(2\pi t)\boldsymbol{\hat{y}}.
\end{equation}
The corresponding bulk fluid excursion is 
\begin{equation}\label{eq:bulkexcursion}
    \boldsymbol{x}_b = -A\cos(2\pi t)\boldsymbol{\hat{y}},
\end{equation}
where $A$ appears due to the differences in nondimensionalization of velocities and length scales, according to Eq.~\eqref{eq:nondim}. An equivalent derivation for the excursion of the flow is given in Appendix~\ref{analyticalvelocityprofiles}.

The fluid phase is governed by the continuity equation for an incompressible fluid
\begin{equation}\label{eq:continuity}
    \boldsymbol{\nabla\cdot u} = 0,
\end{equation}
and the Navier-Stokes equation for a Newtonian fluid
\begin{equation}\label{eq:navierstokes}
    \frac{1}{2\pi}\frac{\partial \boldsymbol{u}}{\partial t} + A\left(\boldsymbol{u\cdot\nabla}\right)\boldsymbol{u} = -\boldsymbol{\nabla} p + \frac{1}{2}\delta^2\nabla^2\boldsymbol{u} + \cos\left(2\pi t\right)\hat{\boldsymbol{y}},
\end{equation}
both already in dimensionless form. Note that the external pressure gradient [Eq.~\eqref{eq:externalpressure}] is written explicitly in the last term. The first term on the right-hand side only contains the pressure gradient due to the flow around the spheres and near the plates.

Halfway between the plates, the stress-free boundary condition
\begin{equation}\label{eq:bc_symmetry}
    \left.\frac{\partial u}{\partial z}\right|_{z=H/2} = 0, \quad
    \left.\frac{\partial v}{\partial z}\right|_{z=H/2} = 0, \quad
    \left.w\right|_{z=H/2} = 0,
\end{equation}
is enforced to reduce the size of the computational domain. This boundary condition acts as a symmetry plane, such that the computational domain still describes an oscillating channel flow between two plates. Formally, this symmetry plane adds virtual particles that touch the top plate. The effect of these virtual particles can be neglected if $H\gg1$.

The description up to this point is valid for either an oscillating box or an oscillating channel flow. The boundary condition at the bottom of the domain discriminates between the two systems. It is given by 
\begin{equation}\label{eq:bc_bottom}
    \left.\boldsymbol{u}\right|_{z=0} = \left\{
    \begin{array}{ll}
      \sin(2\pi t)\boldsymbol{\hat{y}}, & \text{(oscillating box)} \\[8pt]
      \boldsymbol{0}, & \text{(oscillating channel flow)}
    \end{array} \right.
\end{equation}
such that the no-slip/no-penetration bottom either moves in unison with the bulk flow or is fixed in space. The latter condition introduces a shear in the velocity field, which leads to the formation of a Stokes boundary layer in the region where the viscous forces balance the driving oscillating pressure gradient. The (dimensionless) thickness of the Stokes boundary layer is equal to the viscous length scale $\delta$.

\begin{figure*}
    \includegraphics[width=\textwidth]{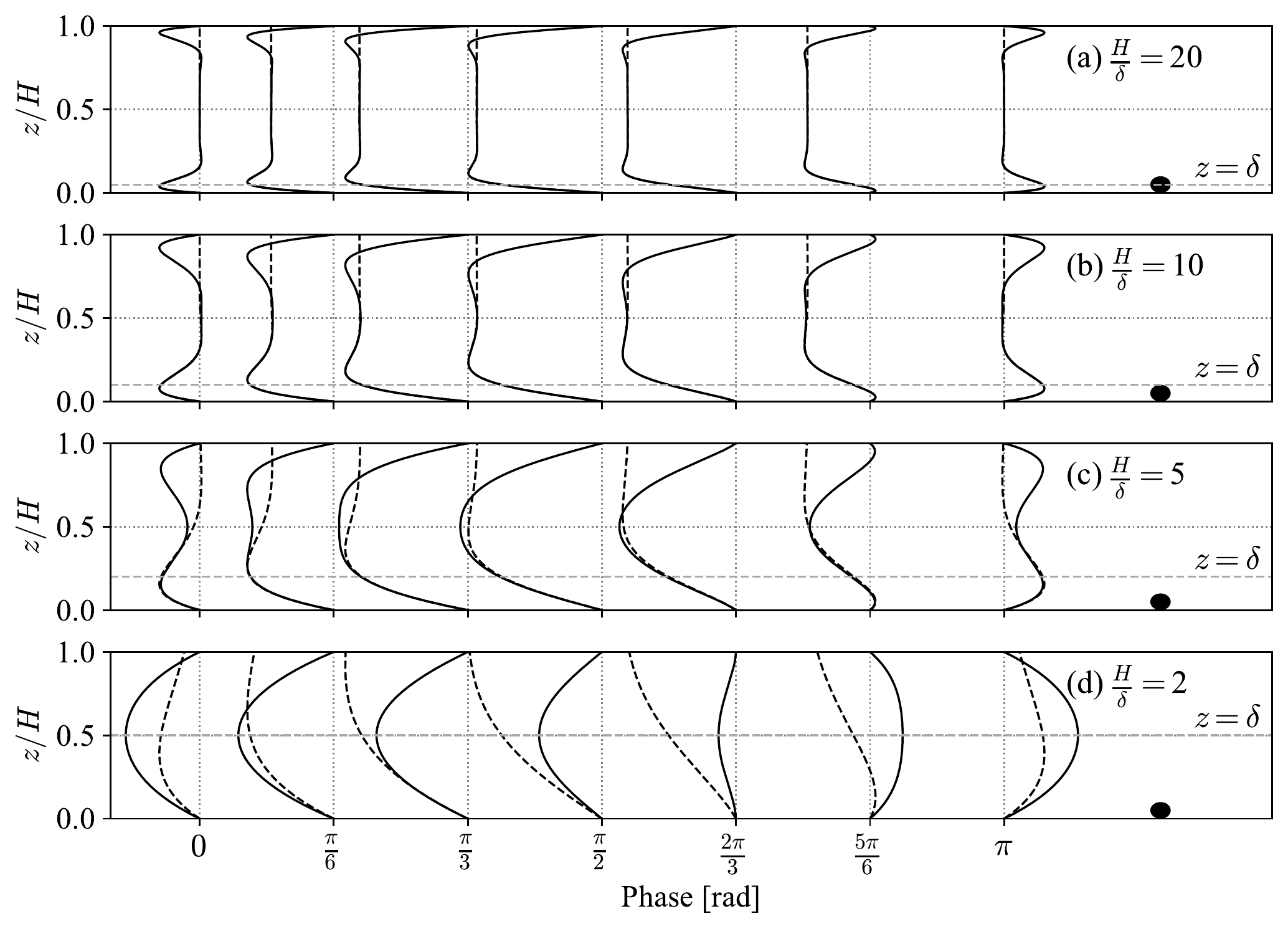}
    \caption{Comparison of laminar velocity profiles of an oscillating flow in a channel (solid) and in an infinitely deep domain (dashed). The profiles are shown side-by-side at different phases, where the dotted lines indicate zero velocity. For large values of $H/\delta$, the top and bottom boundary layers do not interact, and the flow near the bottom of the channel rapidly converges to that of the deep domain. In addition, the black circle represents the particle size used in this study ($D=H'/10$) and the horizontal dashed line indicates the boundary layer thickness, i.e. where $z=\delta$.}
    \label{fig:figure1}
\end{figure*}

If the height of the domain is sufficiently large compared to the boundary layer, i.e. $H/\delta \gg 1$, there is no overlap of top and bottom boundary layers. In such a case, the bottom half of the oscillating channel flow should be equivalent to that of an infinitely deep domain.
To confirm this, Fig.~\ref{fig:figure1} shows vertical profiles of the horizontal velocity at different phases and for four values of $H/\delta$. The analytical equations describing these profiles are given in Appendix~\ref{analyticalvelocityprofiles}. When $H/\delta \gtrsim 5$, the solution for the oscillating channel flow rapidly converges to that of the infinitely deep domain, and the two become equivalent from the perspective of the particles. Hence, in this case, the results hold for both oscillating channel flows and oscillatory flows over a solid plane wall in an infinitely deep domain.

\subsection{Particle motion}\label{sec:formulation_particle} 
\subsubsection{Governing equations}
The particle motion is also presented in nondimensional form. Using the same typical scales ($D$, $2\pi/\omega$, and $A'\omega$) as in Eq.~\eqref{eq:nondim}, we introduce the additional dimensionless variables:
\begin{equation}\label{eq:nondim2}
    \boldsymbol{u}_s = \frac{\boldsymbol{u}_s'}{A'\omega}, \quad
    \boldsymbol{\tau} = \frac{D\boldsymbol{\tau}'}{\rho_f\nu A'\omega}, \quad
    \boldsymbol{\omega}_s= \frac{D\boldsymbol{\omega}_s'}{A'\omega}, \quad
    \boldsymbol{r}=\frac{\boldsymbol{r}'}{D},
\end{equation}
with the particle velocity $\boldsymbol{u}_s'$, the stress tensor $\boldsymbol{\tau}'$, the particle's angular velocity $\boldsymbol{\omega}_s'$, and the vector $\boldsymbol{r}'$ going from the particle's centroid to its surface.

We assume that the gravitational force on the particles is canceled by the sum of the normal force, the lift force, and the buoyant force. Consequently, there is no vertical motion, such that the particle motion is restricted to a two-dimensional (2D) horizontal plane and the particles are always in contact with the bottom. Nonetheless, the particles are free to rotate around any (3D) axis due to e.g. gradients in the flow velocity. The motion of the particles is governed by Newton's laws of motion
\begin{equation}\label{eq:particle_trans_dimless}
    \frac{d\boldsymbol{u}_s}{dt} = 6\frac{\delta^2}{s}\oint\boldsymbol{\tau\cdot}\hat{\boldsymbol{n}}dS +\frac{2\pi}{s}\cos(2\pi t)\hat{\boldsymbol{y}} + 2\pi\left(\frac{s-1}{s}\right)\frac{\mu_c}{\Gamma}\hat{\boldsymbol{f}},
\end{equation}
\begin{equation}\label{eq:particle_rot_dimless}
    \frac{d\boldsymbol{\omega}_s}{dt} = 60\frac{\delta^2 }{s}\oint\boldsymbol{r}\times\left(\boldsymbol{\tau\cdot}\hat{\boldsymbol{n}}\right)dS+20\pi\left(\frac{s-1}{s}\right)\frac{\mu_c}{\Gamma}\boldsymbol{r}\times\hat{\boldsymbol{f}},
\end{equation}  
both in dimensionless form, where $s=\rho_s/\rho_f$ is the particle-fluid density ratio, $\hat{\boldsymbol{n}}$ is the outward vector normal to the surface $S$ of the spherical particle, $\Gamma=A'\omega^2/g$ is the ratio between oscillatory and gravitational acceleration, and $\hat{\boldsymbol{f}}$ is the unit vector that accounts for the relative velocity difference between the bottom plate and particle, as described in more detail by \citet{VanOverveld2022}.

From here on, for the sake of simplicity, we consider that the friction between the particle and the bottom can be neglected, i.e. the Coulomb friction coefficient $\mu_c=0$. In most environmental settings, like for rolling-grain ripples, this assumption is not valid \citep{Mazzuoli2016}. However, \citet{VanOverveld2022} found that simulations where friction is neglected have good agreement with experimental data of an oscillating box at high frequencies\citep{Klotsa2007}.

In absence of particle-bottom friction, the last term in each of Eqs.~\eqref{eq:particle_trans_dimless}~and~\eqref{eq:particle_rot_dimless} is equal to zero, such that the particle motion is independent of $\Gamma$. In combination with the equations for the fluid motion (Eqs.~\eqref{eq:continuity}~and~\eqref{eq:navierstokes}), the full system is uniquely defined by three dimensionless control parameters: $A$, $\delta$ and $s$.

\subsubsection{Relative motion}
Under the assumption that viscous effects are important, i.e. when the Reynolds number is not too large ($\lesssim100$), the stress tensor oscillates harmonically over time. According to Eq.~\eqref{eq:particle_trans_dimless}, the particle translation ($\boldsymbol{x}_s$) in the streamwise direction should then also be sinusoidal, following 
\begin{equation}\label{eq:particle_streamwise}
    \boldsymbol{x}_s = -A_s\cos(2\pi t+\phi)\hat{\boldsymbol{y}},
\end{equation}
where the (dimensionless) excursion length $A_s$ and phase lag $\phi$ are unknown functions of $A$, $\delta$, and $s$. 

Because both the streamwise particle and bulk fluid motion are sinusoidal, so is their relative motion \citep{Wunenburger2002}. The corresponding relative excursion length is
\begin{equation}\label{eq:amplitudes}
    A_r = \sqrt{A^2 + A_s^2 - 2AA_s\cos\left(\phi\right)},
\end{equation}
which follows directly from Eqs.~\eqref{eq:bulkexcursion}~and~\eqref{eq:particle_streamwise}, and is explained in more detail by \citet{VanOverveld2022}.
Similar to $A_s$, $A_r$ is an unknown function of $A$, $\delta$, and $s$, which cannot be set a priori. For $s=1$ in the oscillating box, the particle and fluid are subjected to the same force of acceleration (per unit volume), such that the particle moves in unison with the fluid and $A_r=0$. For $s\rightarrow\infty$, the particle remains stationary in the lab frame, such that $A_r=A$. For $s=7.5$ and $A \lesssim 10$, the empirical scaling $A_r\sim A/\delta^{0.5}$ was found by \citet{VanOverveld2022}.

\citet{Klotsa2007} used $A_r$ instead of $A$ and $s$ to describe the mean state of the oscillating box. Later, \citet{VanOverveld2022} confirmed that only $\delta$ and $A_r$ are important for the generation of the steady streaming flow and its subsequent interaction with the particles. All dependency of the mean equilibrium state on $A$ and $s$ is implicitly incorporated in $A_r$. By replacing the known parameters $A$ and $s$ with the a priori unknown parameter $A_r$, the set of dimensionless quantities that describe the mean equilibrium state of the system is reduced from three ($A$, $\delta$, $s$) to two ($A_r$, $\delta$). 

Conversely, for the oscillating channel flow, we expect that the relative excursion $A_r$ is not a useful quantity for all flow conditions, due to the presence of a Stokes boundary layer above the bottom. While $A_r$ relates the particle motion to the bulk flow, the streamwise particle motion itself is a result of the local, non-uniform flow. We would expect that, when $\delta\gtrsim0.5$, the Stokes boundary layer is sufficiently thick such that the particle feels a non-uniform velocity profile with an average magnitude significantly lower than that of the bulk flow, as illustrated in Fig.~\ref{fig:figure1}. 

We therefore propose a different relative excursion length $A_R$ that takes into account both changes in amplitude and phase of the flow due to the Stokes boundary layer. We consider the relative motion between the particle and the undisturbed flow at the particle center, i.e. at $z=1/2$. The undisturbed oscillatory flow over a fixed plane boundary is given by the analytical expression in Eq.~\eqref{amplitude_analytical_unbounded_particleheight} in Appendix~\ref{analyticalvelocityprofiles}. The relative motion is again a sinusoidal function, now with amplitude $A_R$ and phase lag $\phi_R$:
\begin{eqnarray}\label{eq:AR}
     -A_s\cos(2\pi t+\phi) &+& A\left[\cos(2\pi t)-e^{-1/2\delta}\cos{\left(2\pi t - \frac{1}{2\delta}\right)}\right] \nonumber\\
    &\equiv& A_R\cos(2\pi t + \phi_R).
\end{eqnarray}
We can determine $A_R$ from Eq.~\eqref{eq:AR}, which, with help of Eq.~\eqref{eq:amplitudes}, can be written as
\begin{eqnarray}\label{eq:AR2}
    A_R^2 = A_r^2 + A^2\left[e^{-1/\delta}-2e^{-1/2\delta}\cos\left(\frac{1}{2\delta}\right)\right]\nonumber\\
    +2AA_se^{-1/2\delta}\cos\left(\phi+\frac{1}{2\delta}\right).
\end{eqnarray}
From this expression it follows that $A_R$ depends on $A$, $\delta$, and (implicitly) on $s$. When either $A$ or $\delta$ is varied, $A_R$ can be kept constant as long as $s$ is co-varied.
Moreover, based on Eq.~\eqref{eq:AR2}, we expect the largest deviation of $A_R$ from $A_r$ when $\delta$ is large. Contrarily, in the limit of $\delta\rightarrow0$, the Stokes boundary layer becomes infinitely thin, such that the flow conditions at $z=1/2$ are equal to those in the bulk. In this limit, the oscillating channel flow becomes equivalent to the oscillating box, such that $A_R=A_r$. 

We have previously hypothesized that the oscillating channel flow has an additional degree of freedom compared to the oscillating box: the relative excursion between the fluid and the boundaries. Filling in $A_s=0$ in Eq.~\eqref{eq:AR2} yields the relative excursion between the undisturbed flow at the particle center and the wall:
\begin{eqnarray}\label{eq:ARwall}
    A_{R,\mathrm{wall}} = A\sqrt{1+e^{-1/\delta}-2e^{-1/2\delta}\cos\left(\frac{1}{2\delta}\right)}.
\end{eqnarray}
This quantity is related to the typical shear rate to which the particle is exposed and is only a function of $A$ and $\delta$. So, when $A_R$ is kept constant, by co-varying $A$ and $s$, the shear at the position of the particle changes due to the variation in $A$. So, in the oscillating channel flow, $s$ can not be varied without changing $A_R$ or the typical shear rate.
Contrarily, for the oscillating box, the walls and bulk fluid move in unison, such that $A_{R,\mathrm{wall}}=0$. The aforementioned shear is thus absent. When $A$ and $s$ are now co-varied, such that $A_R$ is kept constant, the relative fluid motion around the particle is also constant. So, $s$ is not a relevant parameter for the oscillating box, as its variation leads to the same relative flow around the particles.

As a consequence of the additional degree of freedom in the oscillating channel flow, the number of dimensionless quantities cannot be reduced, and three quantities ($A_R$, $\delta$, $s$) are needed to describe the system. The local flow conditions around the particles are described by $A_R$ and $\delta$. Therefore, a comparison of the oscillating box and the oscillating channel flow at constant values of $A_R$ and $\delta$ implies that the local flow conditions around the particles are similar in both systems. The extra degree of freedom can then be explored through variation of $s$. Note that changing the value of $s$, while keeping $A_R$ and $\delta$ constant, implies that the value of $A$ also changes accordingly.

\subsection{Numerical method}
The code used in this study is identical to the one used by \citet{VanOverveld2022}, and a related version was recently also used by \citet{Shajahan2020}. Moreover, it is similar to the one used by \citet{Mazzuoli2016}. These sources contain a more extensive description and may be useful to the interested reader.
The fluid phase is solved in the whole domain on a uniform Cartesian grid with spacing $D/16$, such that flow structures can be resolved on a sub-particle level. This resolution is similar to those used by e.g. \citet{Mazzuoli2016} and \citet{Klotsa2007}, and it was previously determined to be sufficient to capture the particle dynamics \citep{VanOverveld2022}. 
Periodic boundary conditions are used in the $x$ and $y$\nobreakdash-directions, with a domain size of $L_x\times L_y = 15\times20$ particle diameters. This size is chosen to keep computational costs relatively low while minimizing the effects of periodic boundary conditions. For the majority of simulations ($\gtrsim90\%$), the bulk fluid excursion length $A$ is smaller than half the domain length, i.e. $A<10$. For the simulations where $A>10$, we have elongated the domain to $L_x\times L_y = 15\times40$ to guarantee that there is no overlap of the wakes (with approximate length $A$, see \citet{VanOverveld2022}) upstream and downstream of the particles. In other words, the interaction of particles with their own wakes through the periodic boundaries is minimized. We have verified that for $A\approx10$, the difference between the two domain sizes in equilibrium is minimal: derived quantities such as, for example, $A_R$ and the mean gap between the particles, are affected less than 1\%. This difference is sufficiently small such that it does not affect the conclusions of this study.  
For the $z$\nobreakdash-direction, the no-slip boundary condition Eq.~\eqref{eq:bc_symmetry} is enforced at $z=0$. The stress-free boundary condition Eq.~\eqref{eq:bc_bottom} is enforced at $z=5$, such that effectively $H=10$. 

Each particle is represented by $746$ points distributed over a spherical shell with a fixed position relative to the centroid. At each point on the shell, a force is added to the fluid such that the local flow and surface velocities match. This is done according to the second-order accurate immersed boundary method (IBM) by \citet{Breugem2012}. 

The dynamics of both the fluid and particles are obtained by integrating Eqs.~\eqref{eq:navierstokes}, \eqref{eq:particle_trans_dimless} and \eqref{eq:particle_rot_dimless} over time using an explicit three-step Runge-Kutta scheme \citep{Wesseling2001}, embedded in a pressure-correction scheme. The time step $\Delta t$ for each simulation satisfies the von Neumann stability criterion \citep{Breugem2012}. Additional restrictions are added to the time step, to ensure that each oscillation is fully and symmetrically resolved. The total number of time steps per oscillation period, $1/\Delta t$, is an even integer.

The interaction between the particles and the bottom is accounted for by a soft-sphere collision model, based on a spring-damper model \citep{Costa2015}. The same model is used for particle-particle collisions, but these are anyway absent in our simulations.
Manual input is required for the dry coefficients of restitution in the direction normal $e_n$ and tangential $e_t$ to the collision. The exact values of these coefficients is likely irrelevant because particle-bottom friction is neglected in our simulations ($\mu_c=0$), and because we are primarily interested in cases where particles are always in contact with the bottom. Nonetheless, we set the values to $e_n=0.97$ and $e_t=0.39$, which are previously used to describe an oblique particle-wall collision between glass materials \citep{Costa2015}. Similar values are used by \citet{Shajahan2020} and \citet{Mazzuoli2016}.
Additionally, a lubrication correction model is used to resolve forces on particles at positions where the space between the particle and the bottom is smaller than the grid size. We refer to the work of \citet{Costa2015} for more details.

At the start of each simulation, two particles are initialized on the bottom in a side-by-side configuration such that the line between their centroids is perpendicular to the oscillation direction. The initial distance between them is varied per simulation since it should be close to the equilibrium distance to save computational costs.

The specific parameter values used in our simulations are given below. A concise overview is given in table~\ref{tab:literatureoverview}, where additionally an indication of the parameter values used in other relevant studies is given. In our simulations, we consider values of $\delta$ equal to $1/1.0$, $1/1.25$, $1/1.5, 1/2.25, 1/3.25, 1/4.5$ and $1/5.5$, here given as reciprocals because $1/\delta$ is set in the code. 
The simulations for the oscillating box with $1/5.5\leq\delta\leq1/1.5$ are the same as used by \citet{VanOverveld2022}. The largest two values, $\delta=1/1.25$ and $1/1.0$, have been added to extend the parameter space.
In the next sections, we refer to the (approximate) decimal form of $\delta$, since it allows for a more straightforward comparison between simulations. 
In dimensionful numbers, a value of $\delta=1.0$ (i.e. the Stokes boundary layer thickness equal to the particle diameter) could correspond to sediment grains with a diameter of $\SI{800}{\mu m}$ (coarse sand) or $\SI{400}{\mu m}$ (medium sand), submerged in water and forced at a frequency of $\SI{0.50}{Hz}$ or $\SI{2.0}{Hz}$, respectively \citep{vanrijn1993}.
We start with the density ratio $s=7.50$, which is identical to that used by \citet{Klotsa2009}, \citet{Klotsa2007}, and \citet{VanOverveld2022}, and similar to $s=7.8$ used by \citet{Wunenburger2002}. Later, when the effects of $A_r$ and $A$ are investigated separately, $s$ is lowered up to $2.65$, which is used by \citet{Mazzuoli2016} to simulate sediment grains.
The excursion length of the bulk flow $A$ is varied between, approximately, $0.37$ and $21.2$.
The corresponding values of the Reynolds number of the oscillatory boundary layer, $\mathrm{Re}_\delta=A'\omega\delta'/\nu=2A/\delta$, are always below 100. This is well below the onset of intermittent or turbulent regimes, which occur around $\mathrm{Re}_\delta\sim O\left(10^3\right)$\,\citep{Kaptein2019}. Our results thus always correspond to the regime where the flow is laminar, whilst exhibiting non-linear effects through, for example, steady streaming flows.

\begin{table*}
    \caption{\label{tab:literatureoverview}Range of values of dimensionless numbers considered in relevant previous studies and in this study. The parameter $A_R$ represents the particle excursion length relative to the undisturbed flow at the particle center, as defined in Eq.~\eqref{eq:AR2}. Note that for the oscillating box, $A_R=A_r$.}
    \begin{ruledtabular}
    \begin{tabular}{llcccc}
                                &   System type   & $s$         & $\delta$    & $A$         & $A_R$        \\
        \citet{Klotsa2007}      & Oscillating box  & $6.9-7.5$   & $0.11-0.45$ & $0.20-6.6$  & $0.073-3.0$  \\
        \citet{VanOverveld2022} & Oscillating box  & $2.65-7.50$       & $0.18-0.67$ & $0.14-16.2$ & $0.245-9.2$  \\
        \citet{Mazzuoli2016}    & Oscillating channel flow  & $2.46-2.65$ & $0.52-1.89$ & $12.1-24.8$ & $\approx5.4$ \\
        This study              & Both             & $2.65-7.50$  & $0.18-1.00$  & $0.37-21.2$ & $0.18-9.2$   \\
    \end{tabular}
    \end{ruledtabular}
\end{table*}

\section{\label{sec:results}Results on particle dynamics}
In this section, we describe the differences in particle dynamics between the oscillating box and oscillating channel flow. These dynamics are the result of particle-fluid interactions and are thus dependent on the flow around the particles. In particular, the equilibrium configuration of the pair is determined by the averaged steady streaming flow. If these flows are significantly different in both systems, we can then expect differences in the particle motion. 

The steady streaming flows are typically described using vorticity patches and often presented in two-dimensional slices of the domain (see, e.g., the work by \citet{Klotsa2009}). Here, we consider the flow field characteristics in all three coordinate directions. On the one hand, it is found that the vorticity in the horizontal $xy$-plane relates to the equilibrium configuration and dynamics of the particles \citep{VanOverveld2022}. On the other hand, we are interested in the effect of the velocity shear in the vertical $z$-direction on the steady streaming flow. Even though the residual of the Stokes boundary layer itself is zero after averaging over an oscillatory period, it can still affect the non-zero steady streaming flow.

\begin{figure*}
    \begin{subfigure}[b]{0.49\textwidth}
        \includegraphics[width=\textwidth,{trim=300 100 300 300}, clip]{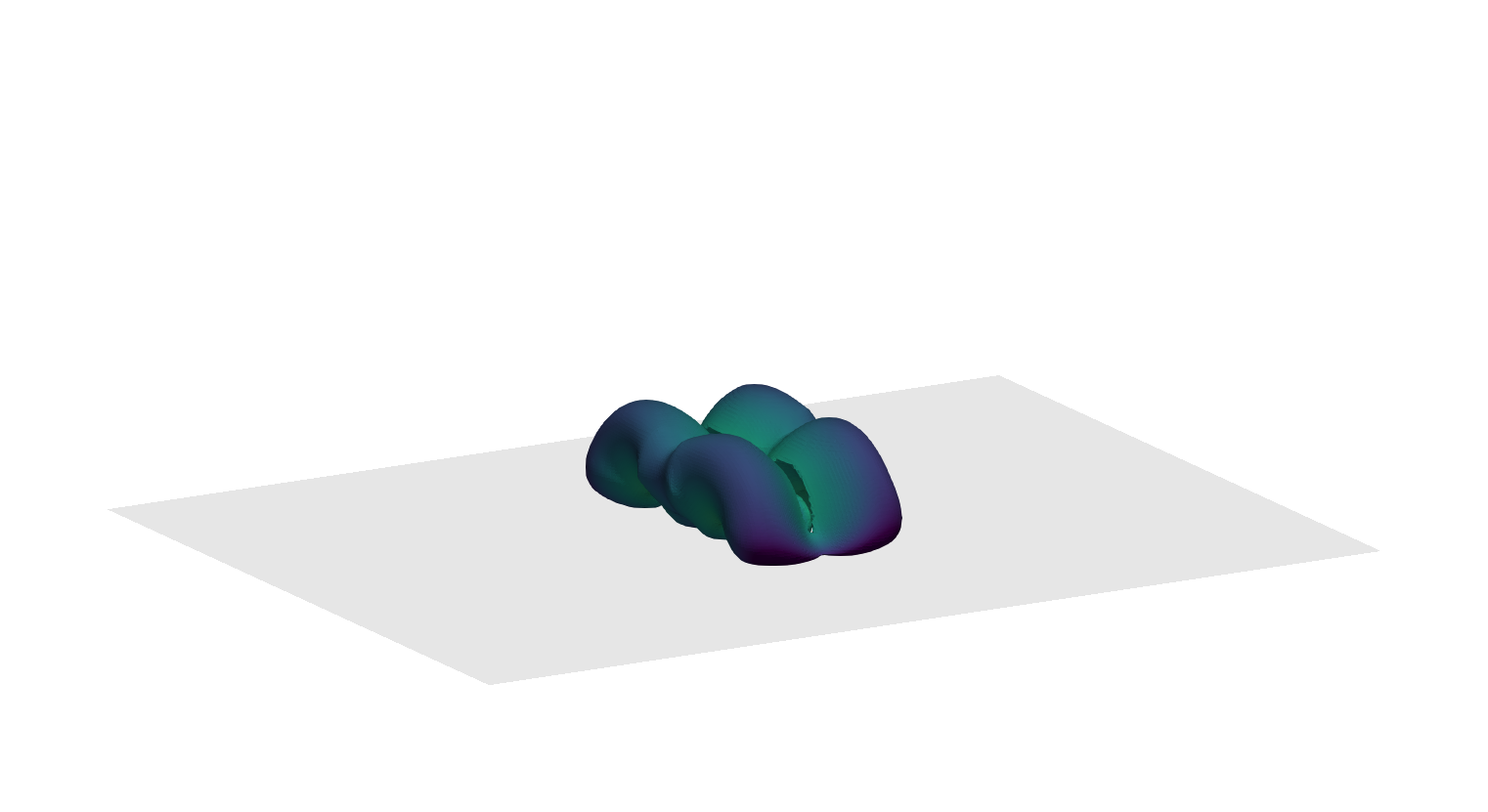}
        \caption{Oscillating box, $A_R\approx1.18$} 
        \label{fig:figure2a}
    \end{subfigure}
    \begin{subfigure}[b]{0.49\textwidth}
        \includegraphics[width=\textwidth,{trim=300 100 300 300}, clip]{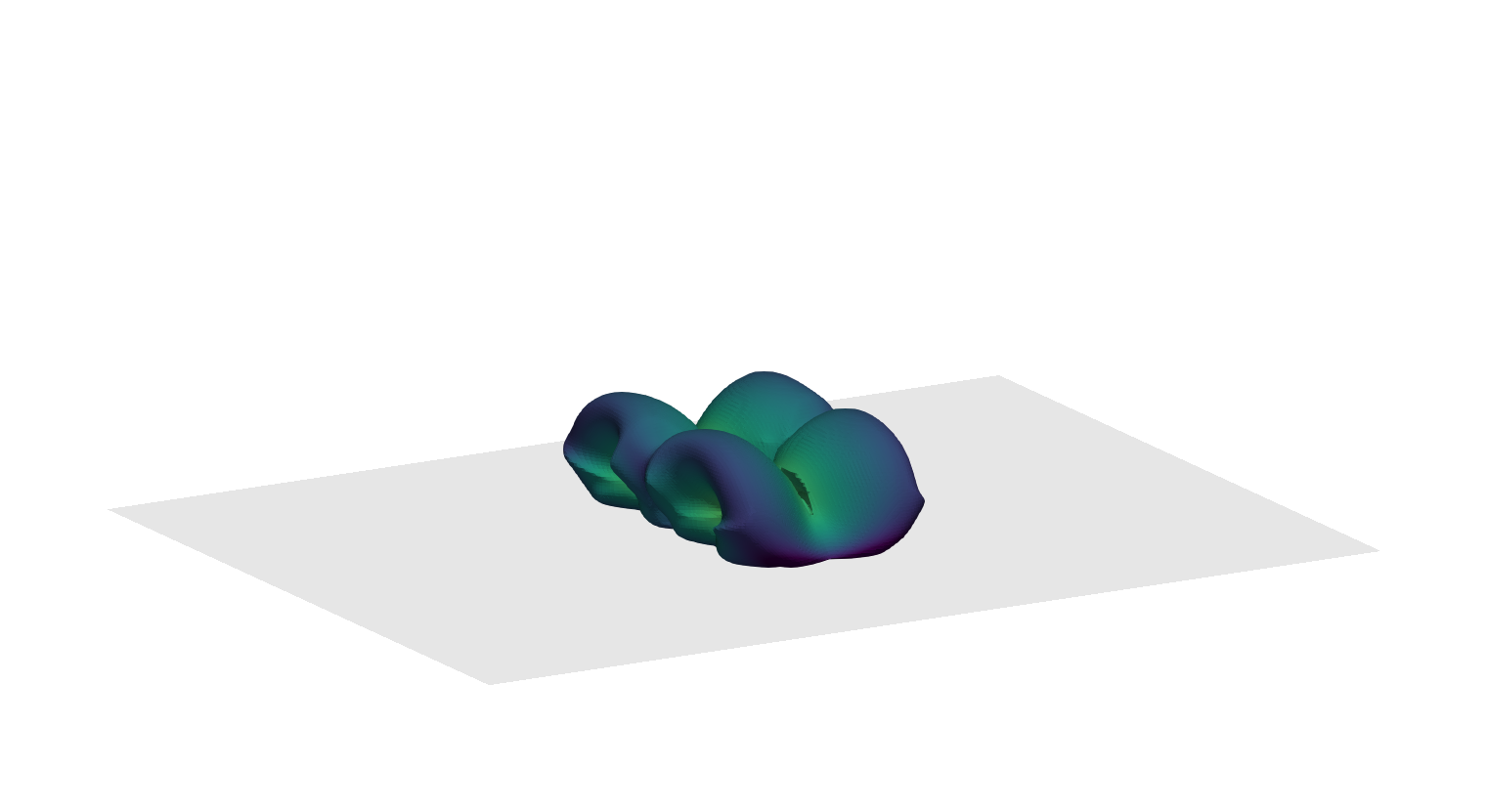}
        \caption{Oscillating channel flow, $A_R\approx1.24$} 
        \label{fig:figure2b}
    \end{subfigure}
        \begin{subfigure}[b]{0.49\textwidth}
        \includegraphics[width=\textwidth,{trim=300 100 300 300}, clip]{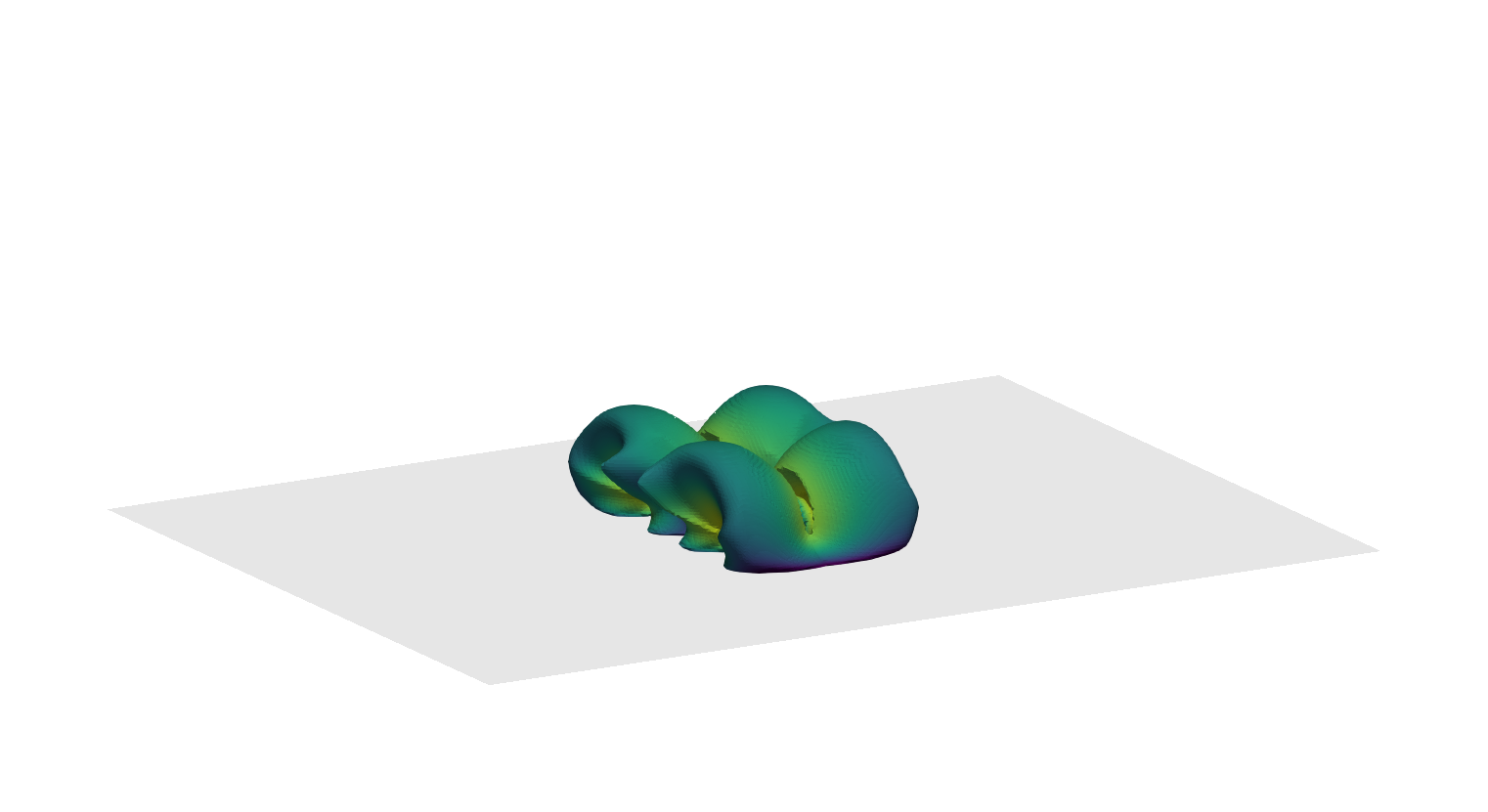}
        \caption{Oscillating box, $A_R\approx2.67$} 
        \label{fig:figure2c}
    \end{subfigure}
    \begin{subfigure}[b]{0.49\textwidth}
        \includegraphics[width=\textwidth,{trim=300 100 300 300}, clip]{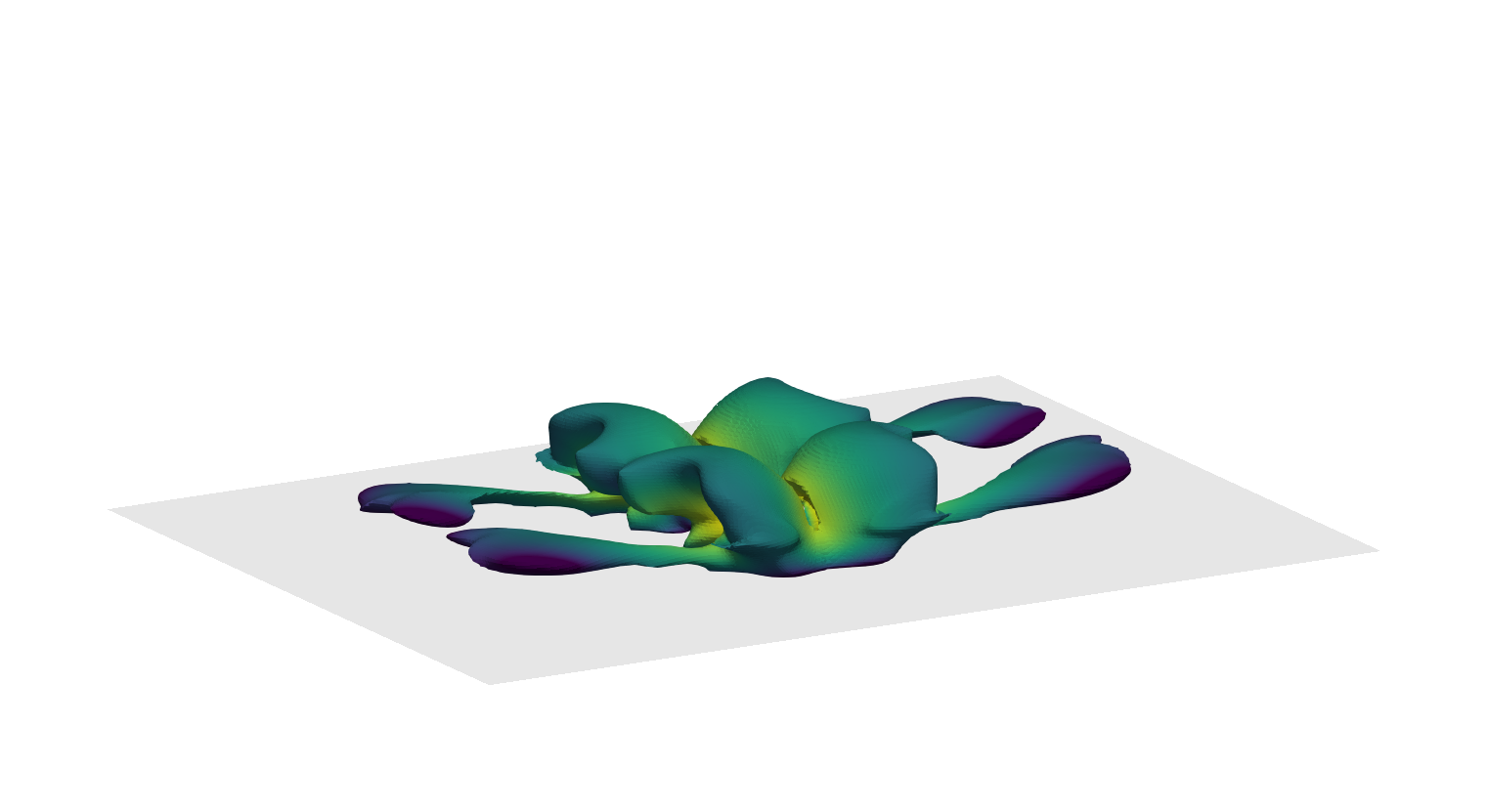}
        \caption{Oscillating channel flow, $A_R\approx2.78$} 
        \label{fig:figure2d}
    \end{subfigure}
    \begin{subfigure}[b]{0.49\textwidth}
        \includegraphics[width=\textwidth]{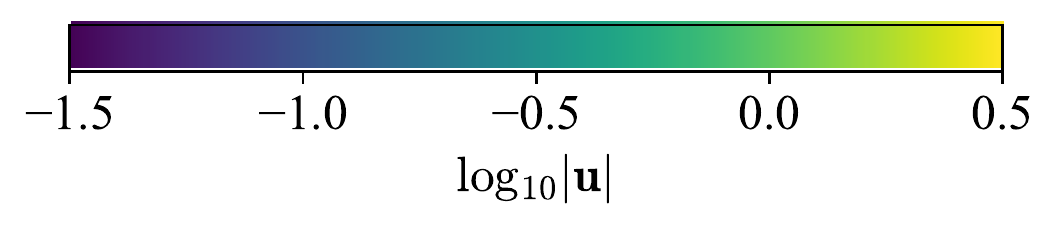}
    \end{subfigure}
    \caption{A three-dimensional view of the oscillation-averaged vortex structures using the $\lambda_2$-criterion\citep{Jeong1995} (isosurfaces of $\lambda_2=-2\times10^{-3}$, chosen slightly below zero for visualization purposes) around the particle pairs for $\delta\approx0.67$ and $s=7.50$. The colors correspond to the logarithm of the velocity magnitude. Animations illustrating the three-dimensionality of the structures are included as supplementary materials.}
    \label{fig:figure2}
\end{figure*}

Specifically, we visualize the three-dimensional vortex structure of the flow, averaged over one oscillation period, using isosurfaces of the $\lambda_2$\nobreakdash-criterion, a method introduced by \citet{Jeong1995}. Figure~\ref{fig:figure2} shows the vortex structures in the oscillating box and oscillating channel flow, for $s=7.50$ and $\delta\approx0.67$. The value of $\delta$ is relatively large to clearly illustrate the effect of the Stokes boundary layer. Two different values of the relative amplitude $A_R$ ($A_R\approx1.2$ and $2.7$) are chosen to allow for a comparison between the oscillating box and oscillating channel flow at similar flow conditions around the particles. These two values of $A_R$ roughly correspond to the viscous- and advection-dominated regime in the oscillating box\citep{VanOverveld2022}, for which, we recall that $A_R=A_r$.

In Fig.~\ref{fig:figure2a}, half of a ring-like vorticity structure is found on the upstream and downstream sides of each particle. These coherent structures correspond to the half vortex rings discussed by \citet{Klotsa2007}. Also for higher $A_R$ values in the oscillating box, shown in Fig.~\ref{fig:figure2c}, do the main vortices stay close to the particles. These structures are stretched in the streamwise direction with respect to the case with lower $A_R$ due to the increase in relative excursion length. Similar structures are found for the oscillating channel flow for a similar $A_R$-value, shown in Fig.~\ref{fig:figure2b}. However, the surfaces are more stretched and elongated diagonally upwards, away from the particles and bottom. The height-dependent stretching of the vortices is a reflection of the vertical gradients in the flow field in the oscillating channel flow: the typical excursion length of both the flow and the vortices increases with distance to the bottom. 

The structures in the oscillating channel flow with $A_R\approx 1.24$ (Fig.~\ref{fig:figure2b}) and in the oscillating box with $A_R\approx 2.67$ (Fig.~\ref{fig:figure2c}) look very similar. However, there is an important difference close to the bottom. There, the streamwise extension of the structures is increased for the oscillating box, while for the oscillating channel flow, it is almost zero. 
For the oscillating channel flow at higher $A_R$, shown in Fig.~\ref{fig:figure2d}, the half-rings around the particles are further elongated in the streamwise direction than for the other three cases, but this elongation remains restricted close to the bottom. In addition, thin `plumes' appear on both streamwise sides of the particles. These plumes are connected to the bottom close to the particle and are angled upwards and away from the particles.

All in all, Fig.~\ref{fig:figure2} illustrates that the averaged flow fields close to the particle pairs are affected by the vertical velocity gradients in the oscillating channel flow, especially for large $A_R$-values. In the rest of this section, we address how the differences in steady streaming flow are related to differences in the equilibrium state of the system. We consider the streamwise particle motion, mean particle separation, and spanwise particle motion, as a function of $A_R$ and $\delta$, while keeping $s$ constant.

\subsection{Streamwise particle motion}\label{results_streamwise}

\begin{figure*}
    \includegraphics[width=\textwidth]{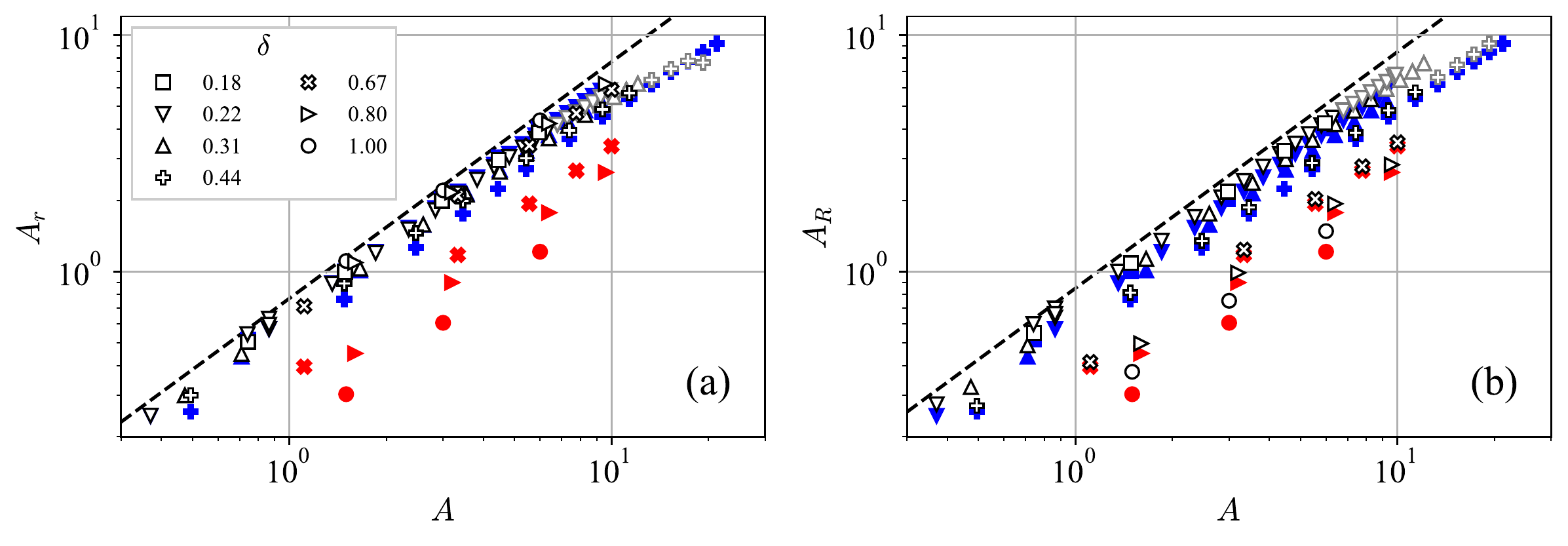}
    \caption{(a) The particle excursion length relative to the bulk flow $A_r$ (given by Eq.~\eqref{eq:amplitudes}) and (b) the particle excursion length relative to the undisturbed flow at the height of the center of the particle $A_R$ (given by Eq.~\eqref{eq:AR2}), both as a function of the absolute excursion length of the bulk flow $A$. For $A_r$, the data of the oscillating channel flow (empty symbols) agree with those of the oscillating box (filled symbols) for $\delta<0.5$ (in blue) but not for $\delta>0.5$ (in red). For $A_R$, the data from both systems agree well. The dashed lines have slopes of 1 dec/dec. Cases where the particles lose contact with the bottom are marked in gray.
    }
    \label{fig:figure3}
\end{figure*}

First, we present the results on the streamwise particle motion as a function of $A$, for $s=7.50$ and different values of $\delta$. Figures~\ref{fig:figure3}(a)~and~(b) show the normalized excursion length of the particle relative to the bulk flow $A_r$ and relative to the undisturbed flow at the particle center $A_R$, respectively. In both cases, we compare the simulations for the oscillating channel flow against those from the oscillating box, which otherwise have identical settings. Note again that $A_R=A_r$ for the oscillating box. 

For both systems and $A_R\lesssim5$, the relative amplitudes are proportional to $A$, as indicated by the dashed lines. Around $A_R\approx5$, this proportionality breaks down, as can be seen by the increasing distance between the symbols and the dashed line. This deviation is due to a superlinear increase of $A_s$ (the absolute particle excursion length, see Eq.~\eqref{eq:particle_streamwise}) with $A$ and is also found in experiments alike to our simulations \citep{Martin1976,Chan1974}.

The behavior of $A_s$ can be understood based on the local, ambient flow around the particle, which is characterized by the particle Reynolds number $\mathrm{Re}_p=A_R'\omega D/\nu=2A_R/\delta^2$. As $A_R$ increases, so does $\mathrm{Re}_p$. When $A_R=5$, the typical values are $\mathrm{Re}_p\approx50$ for $\delta\approx 0.44$, and $\mathrm{Re}_p\approx200$ for $\delta\approx0.22$. 
For these values of $\mathrm{Re}_p$, the drag coefficient becomes larger than what would be expected from Stokes' law (based on uniform flow). For example, the drag coefficient of a sphere in uniform flow at $\mathrm{Re}_p=100$ is about $4.4$ times larger than the value obtained from linear (Stokes) drag \citep{Abraham1970,Flemmer1986}. This non-linear increase in the drag causes the particles to move more with the surrounding flow, and as a result, the relative amplitudes are lower than expected from the linear scaling.

Additionally, at large values of $A$, the particles sometimes lose contact with the bottom. These simulations are indicated with gray symbols in Fig.~\ref{fig:figure3} and subsequent figures. We stress that the assumption of two-dimensional particle motion is violated, and these results should thus not be considered when drawing conclusions using the theoretical framework outlined in section~\ref{sec:formulation_particle}. In some extreme cases, the particles are lifted from the bottom for as much as 50\% of each oscillation period, reaching a maximum height of $0.15$ (equivalent to $15\%$ of the particle diameter). Nonetheless, the vertical particle motion is only found in the oscillating channel flow, which implies that it is due to the vertical velocity gradients. In fact, it is well-known that a shear flow can exert a net lift force on a sphere \citep{dandy1990,asmolov1999}. For low particle Reynolds numbers, \citet{saffman1965} proposed that the (dimensionfull) lift force on a small sphere in a uniform shear flow with shear rate $\dot\gamma$ is equal to $K'\rho_f A'_R\omega D^2\sqrt{\dot\gamma\nu}$, where $K'\approx81.2$. 
This expression can be applied to the oscillating channel flow for large values of $\delta$, i.e. when the Stokes boundary layer resembles a shear flow on the scale of the particle. For the maximum shear rate, we use $\dot\gamma=A\omega/\delta$, which is derived from the analytical velocity profiles in Appendix~\ref{analyticalvelocityprofiles}. The lift force, using the same nondimensionalization as the other forces on the right hand side of Eq.~\eqref{eq:particle_trans_dimless}, is then given by $KA_R\sqrt{\left(\delta/A\right)}/s$, where $K$ is a constant. 

A particularly relevant quantity for the vertical particle motion is the ratio between the upward lift force and the net downward force (gravity minus buoyancy). This ratio is proportional to $A_R\Gamma\sqrt{\left(\delta/A\right)}/(s-1)$. So, for increasing values of $A_R$, and the other parameters kept constant, the lift force becomes larger with respect to gravity. Once it gets sufficiently large to overcome the net downwards force, the particle gets lifted, which happens for the gray markers in Fig.~\ref{fig:figure3}.

The proportionality (i.e. the ratio $A_r/A$) in Fig.~\ref{fig:figure3}(a) clearly depends on $\delta$. For small values of $\delta$ ($\delta\lesssim0.5$, for the oscillating box indicated by blue symbols in Fig.~\ref{fig:figure3}) the values of $A_r$ in both systems are almost identical for a given value of $A$. For example, empty square symbols fall on top of the blue square symbols. In these cases, the Stokes boundary layer is sufficiently thin, such that the particle mainly `feels' the bulk flow. Contrarily, for higher values of $\delta$ ($\delta \gtrsim 0.5$, for the oscillating box indicated by red symbols in Fig.~\ref{fig:figure3}), the values of $A_r$ in the oscillating channel flow clearly differ from that of the oscillating box. In this regime, the Stokes boundary layer is sufficiently thick such that the particle feels a non-uniform velocity profile over most of its height. The discrepancy in the values of $A_r$ then emerges, because the streamwise particle motion is governed by the local, non-uniform ambient flow, whereas $A_r$ relates the particle motion to the bulk flow.

When considering $A_R$, in Fig.~\ref{fig:figure3}(b), the symbols from the oscillating channel flow (empty symbols) agree with those of the oscillating box (filled symbols) for all values of $\delta$ and $A$ considered. This is not a trivial result, because for given values of $\delta$ and $A$, both the ambient (undisturbed) flow and the absolute particle motion are not the same in both systems. Still, the relative motion between the two is such, that $A_R$ has a similar value in both systems. The good agreement between the systems supports our choice to use the same relative excursion length $A_R$ in both.

Due to the good agreement in values of $A_R$ between the data sets in Fig.~\ref{fig:figure3}(b), we expect that $A_R$ is described by the same scaling in both systems. For the oscillating box, the empirical scaling $A_r \sim A/\delta^{0.5}$ was proposed by \citet{VanOverveld2022}. However, this scaling fails to accurately describe the data for the additional, larger $\delta$ values ($\delta=0.8$ and 1.0) considered here. We propose a more general relationship between $A_R$ and $A$ based on theoretical arguments following the analysis of the translation of a small spherical particle in an unbounded oscillating flow at low Reynolds numbers. The trajectory of such a particle is described by the Basset-Boussinesq-Oseen equation, which is analytically solved in Appendix~\ref{BBOmodel}. This yields an expression for the ratio $A_R/A$ such that
\begin{equation}\label{eq:Fsfd_text}
    F(s,f,\delta)\equiv \frac{A_R}{A} = \frac{2(s - 1)}{\sqrt{(9f\delta)^2(2f\delta+1)^2+(9f\delta+2s+1)^2}},
\end{equation}
where $f$ is a unknown scalar that corrects for the presence of the bottom. 

\begin{figure}
    \centering
    \includegraphics[width=0.485\textwidth]{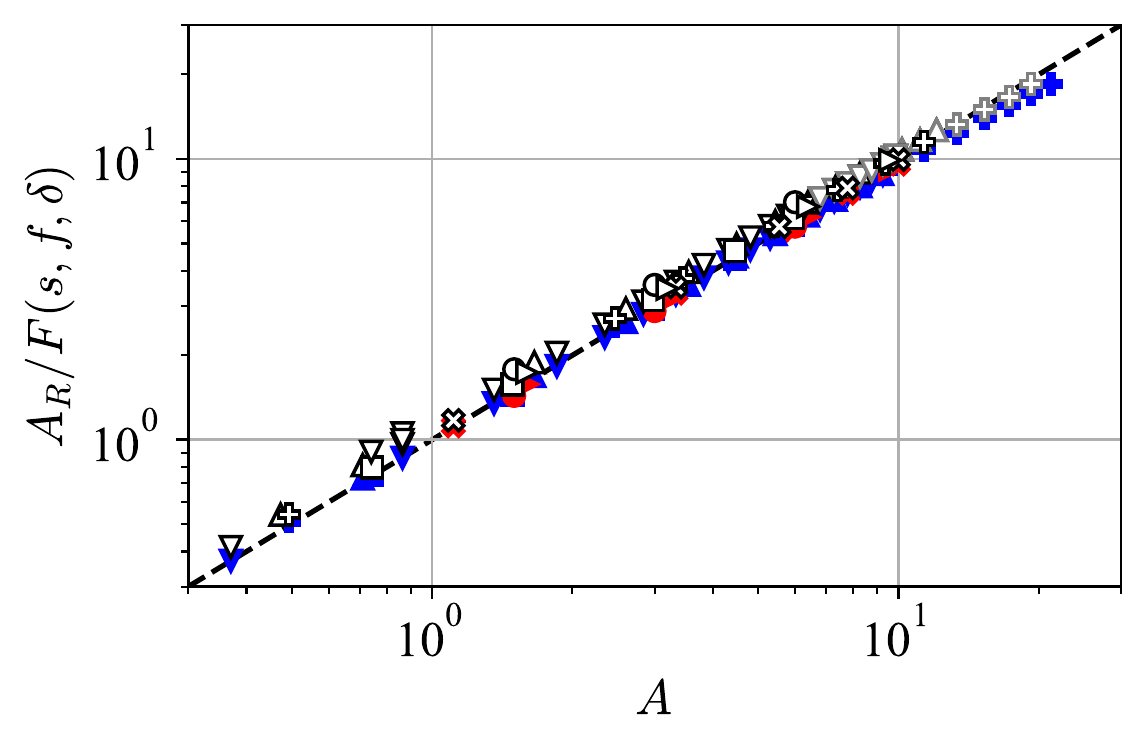}    
    \caption{The particle excursion length relative to the undisturbed flow at center particle height $A_R$, scaled with $F(s,f=1.5,\delta)$ (see Eq.~\ref{eq:Fsfd_text}), as a function of the amplitude of the bulk flow $A$, for $s=7.5$.
    The symbols are identical to those in Fig.~\ref{fig:figure3}. The data from the oscillating channel flow (empty symbols) and the oscillating box (filled symbols) collapse onto the identity line for all values of $\delta$ and $A$ considered, thus $A_R\simeq AF(s,f,\delta)$.
    }
    \label{fig:figure4}
\end{figure}

In Fig.~\ref{fig:figure4}, we have scaled $A_R$ with $1/F(s,f,\delta)$, after which the data collapse onto the identity line for all values of $\delta$ considered. The correction factor $f=1.5$ is empirically determined and implies that the viscous drag on the particles is approximately $1.5$ times larger compared to the drag in an unbounded system. 
Similar values for $f$ have been found for slightly different systems in previous studies, such as the factor $f=1.7$ for the drag on a spherical particle moving close to a wall under influence of a Couette flow \citep{goldman1967slow}. Alternatively, for a sphere moving close to a wall in a quiescent fluid, the drag force scales as $-\ln(w/D)$, with $w$ the gap between particle and wall \citep{goldman1967,oneill1967}. As $w$ tends to zero, the drag diverges to infinity.
We stress that our reported value $f=1.5$ is likely not a universal constant, but that it depends on the interaction between the particle and bottom. In the current numerical method, the separation between the particle and bottom is ill-defined. On the one hand, this is due to the spring-damper model that allows for slight overlap of particle and bottom \citep{Costa2015}. On the other hand, this is due to the non-sharp particle-fluid interfaces in the immersed boundary method used here \citep{Breugem2012}. Nevertheless, this value of $f$ signifies that the bottom plays an important role in the streamwise particle motion, even in the oscillating box.

\subsection{Mean particle separation}\label{results_mean}

\begin{figure*}
    \includegraphics[width=\textwidth]{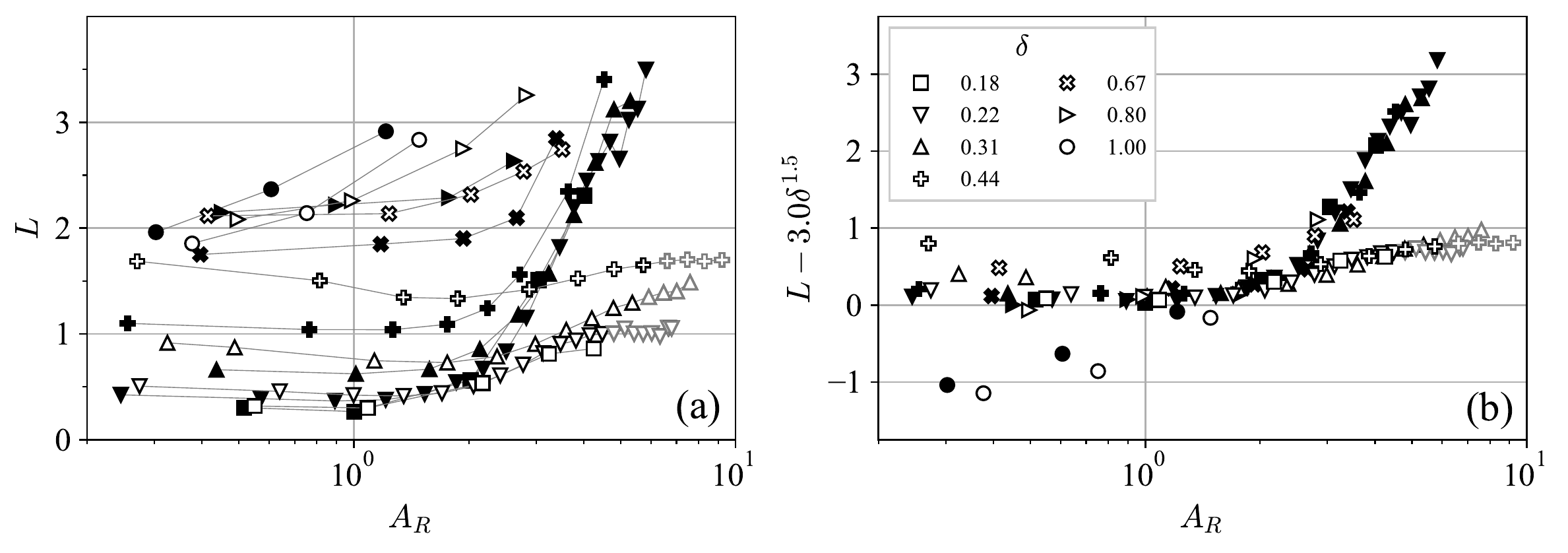}
    \caption{(a) The mean spacing between the particles $L$ as a function of the relative excursion length $A_R$, for $s=7.5$. Lines are added between symbols to guide the eye. (b) The same data is adjusted by subtracting $3.0\delta^{1.5}$, such that the filled symbols, corresponding to the oscillating box, collapse onto a curve. Note the divergence for $A_R\gtrsim2$ between the data from the oscillating box and the data from the oscillating channel flow (empty symbols). Cases where the particles lose contact with the bottom are shown in gray.}
    \label{fig:figure5}
\end{figure*}

The (normalized) mean gap between the particles $L$ as a function of $A_R$ and $\delta$ is shown in Fig.~\ref{fig:figure5}(a) for both systems. For the oscillating box, the relation 
\begin{equation}\label{eq:meangap}
   L\approx 3.0\delta^{1.5}+0.03A_R^3
\end{equation}
holds. This relation gives a transition around $A_R\approx 2$ between a viscous- and an advection-dominated regime \citep{VanOverveld2022}. Below this transition, $3.0\delta^{1.5}$ can be subtracted from $L$ (shown in Fig.~\ref{fig:figure5}(b)), such that most filled symbols collapse onto a single curve for all values of $A_R$. Only the cases with the largest value of $\delta$ ($\delta=1.0$) are an exception to the collapse, with lower-than-expected values of $L$, especially for $A_R<1$. Above the transition, the last term of Eq.~\eqref{eq:meangap} starts to dominate, such that the gap rapidly grows as $L\propto A_R^3$. This significant increase in the mean gap does not occur for the oscillating channel flow. Instead, for low $\delta$ ($\delta\lesssim0.44$), the data (in terms of $L-3.0\delta^{1.5}$) collapse onto a different curve that is weakly dependent on $A_R$ and converges to a plateau at $L-3.0\delta^{1.5}\approx 0.8$. The major difference with the oscillating box is thus the absence of a significant gap increase.

Nonetheless, when $\delta\lesssim0.22$ and $A_R\lesssim2$, the data from both systems (i.e. the empty and filled symbols) show good agreement. This is expected because, in the limit of $\delta\rightarrow0$, the particles feel only the bulk flow and the two systems are equivalent. For small values of $\delta$ ($\delta\lesssim0.22$), the Stokes boundary layer is sufficiently thin, such that it hardly affects the mean gap.

When $\delta$ increases ($\delta\approx0.31$ and $0.44$) the mean gap in the oscillating channel flow gets a weak negative dependence on $A_R$ (see Fig.~\ref{fig:figure5}(a)). The particles are thus drawn closer to each other when their excursion length increases. This phenomenon is not found for the oscillating box and is addressed further in section~\ref{sec:resultsDensityRatio}.

Upon increasing $\delta$ further ($\delta\gtrsim0.67$), for $A_R\lesssim1.0$, the typical values of $L$ increase, but do not become much larger than approximately $2$. The particle interactions are weak in this part of the parameter space, where $A_R$ is small and $\delta$ is large (e.g. for the circles, right-pointing triangles, and crosses in Fig.~\ref{fig:figure5}(a)). As a consequence, the simulations take long to converge: it typically takes hundreds of oscillation periods for the system to reach an equilibrium state.

In fact, due to the high computational costs of the slow-converging simulations, instead of simulating until the system reaches an equilibrium, we extrapolate the numerical particle trajectories to obtain the numerical values of the quantities that describe the equilibrium configuration (e.g. $L$). Details for the fitting functions can be found in Appendix~\ref{fittingequations}. 

\begin{figure*}
    \includegraphics[width=\textwidth]{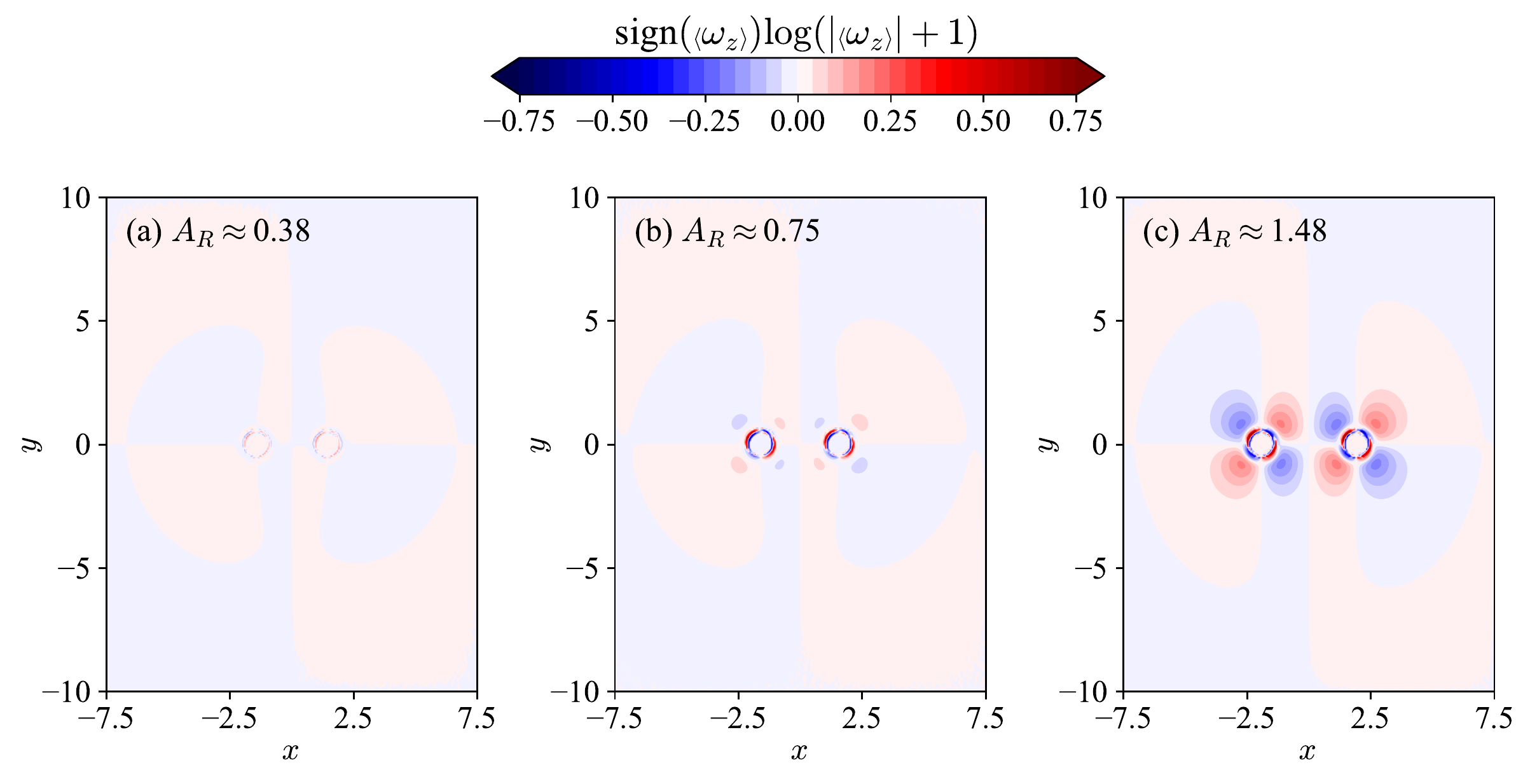}
    \caption{The time-averaged proxy for the vorticity in the $xy$-plane going through the centers of the particles for the oscillating box with $\delta=1.0$, corresponding to the filled circles in Fig.~\ref{fig:figure5}. Note the logarithmic color scale.}
    \label{fig:figure6}
\end{figure*}

The increase in the convergence time is due to the weakness of the steady streaming flow which is responsible for the particle interaction. The weakness can be quantified by defining the time-averaged vorticity in the $xy$\nobreakdash-plane going through the particle centers
\begin{eqnarray}\label{eq:meanvorticity}
    \left<\omega_z\right> &=& \int_{t}^{t+1} \left.\left(\frac{\partial u_y}{\partial x}-\frac{\partial u_x}{\partial y}\right)\right|_{z=1/2}dt' \nonumber\\
    &\approx& \frac{1}{N}\sum\limits_{i=0}^{N} \left.\left(\frac{\partial u_y}{\partial x}-\frac{\partial u_x}{\partial y}\right)\right|_{z=1/2},
\end{eqnarray}
in which the integral is replaced by an average of the flow fields at $N=20$ times within a single oscillation period. In Fig.~\ref{fig:figure6}, the time-averaged vorticity is shown for the three simulations with $\delta=1.00$ (circles in, e.g., Fig.~\ref{fig:figure6}). For the largest excursion length ($A_R\approx1.48$), the vorticity distribution around each particle is qualitatively similar to that found in previous work \citep{Klotsa2007,Jalal2016,VanOverveld2022}. Upon halving the value of $A_R$, the vorticity magnitude in the `outer' patches reduces, such that only a thin layer remains around each particle, corresponding to the particle boundary layer in which most vorticity is produced. When halving $A_R$ again, the vorticity diminishes to almost zero in the whole plane, such that there is nearly no steady streaming flow. For the other values of $\delta$, a similar decrease of the vorticity is found, but not as drastic as shown in Fig.~\ref{fig:figure6}. 

\begin{figure}
    \includegraphics[width=0.48\textwidth]{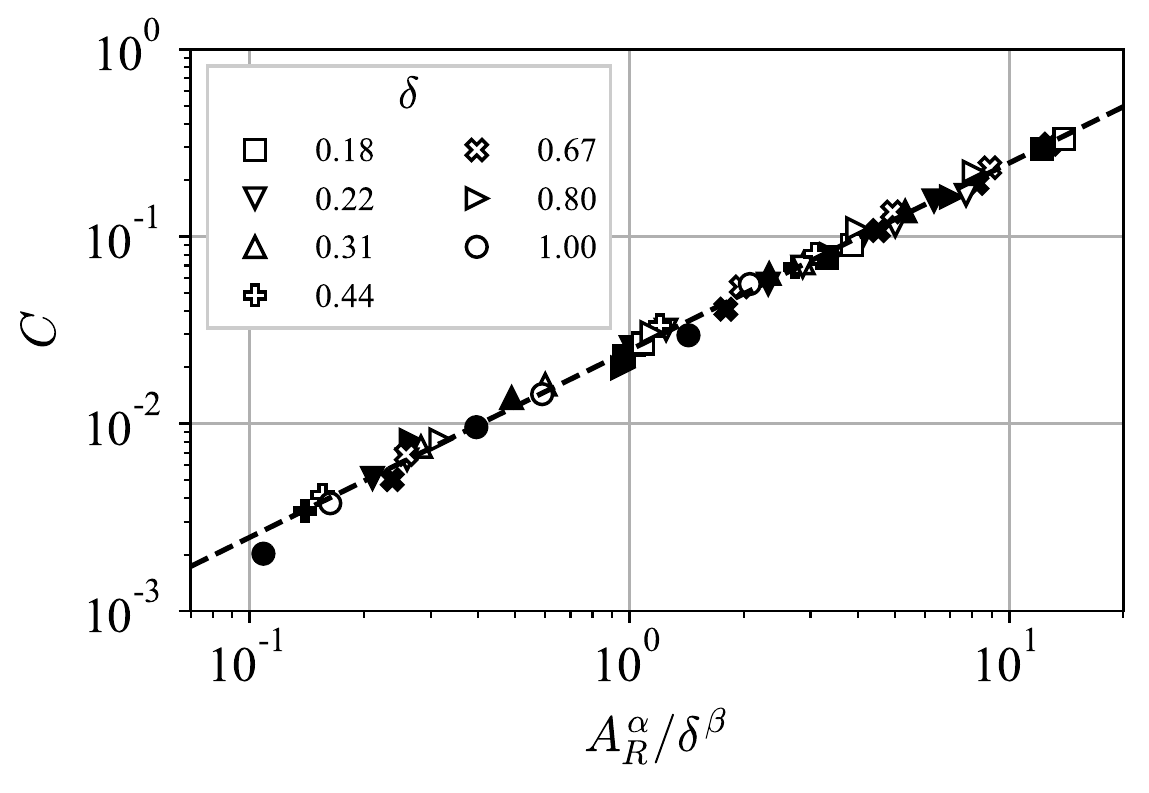}
    \caption{The space-time-averaged vorticity $C$ (see Eq.~\eqref{eq:avgmeanvorticity}) as a function of $A_R^\alpha/\delta^{\beta}$. Least squares analysis on the logarithmic values yields $\alpha\approx1.9\pm0.2$ and $\beta\approx0.7\pm0.2$ for the data to collapse onto a line. The dashed line has a slope of 1 dec/dec.}
    \label{fig:figure7}
\end{figure}

We estimate the total strength of the steady streaming flow using the spatial average of the absolute value of the time-averaged vorticity, defined as 
\begin{equation}\label{eq:avgmeanvorticity}
    C \equiv \frac{1}{L_x L_y}\iint \left|\left<\omega_z\right>\right| dxdy.
\end{equation}
The value of $C$ is shown in Fig.~\ref{fig:figure7} as a function of $A_R^\alpha/\delta^\beta$, where $\alpha$ and $\beta$ are fitting exponents computed so that the data collapse onto a line with a slope of 1 dec/dec. We consider only simulations with $A_R\lesssim2$, to focus on the simulations for which the time-averaged vorticity diminishes to almost zero, as in Fig.~\ref{fig:figure6}. We obtain that 
\begin{equation}\label{eq:C_AR_delta}
    C\sim A_R^{1.9}/\delta^{0.7}.
\end{equation} 
The data collapse suggests that the total steady streaming flow is weak when either viscous dissipation is strong (large $\delta$) or the production of vorticity is weak (small $A_R$). In either case, the flow field approaches the Stokes regime in which non-linear effects do not play a role. Indeed, the cases for which the steady streaming flow is weak or almost absent correspond to the lowest values of $C$ (typically, $C\lesssim 5\times10^{-2}$).
Nonetheless, the simulations with such low values of $C$ should be considered with care since most have not reached an equilibrium configuration due to the long convergence times. It is mainly for these simulations that the extrapolation of the numerical particle trajectories needs to be performed (as mentioned before with an approach discussed in Appendix~\ref{fittingequations}).

For the simulations in Fig.~\ref{fig:figure7}, the particle Reynolds number $\mathrm{Re}_p=2A_R/\delta^2$ varies between $0.7$ and $10$, i.e. with a spread of more than an order of magnitude. The relatively low values of the particle Reynolds number indicate that viscous effects are important. However, $\mathrm{Re}_p$ is not the dimensionless parameter that determines the value of $C$. 

In addition to the physical limitations, the current numerical method becomes more expensive as $C$ becomes smaller, since the number of time steps per oscillation period needs to increase rapidly to account for the increasing viscous dissipation \citep{Breugem2012}. The combination of smaller time steps and longer convergence times severely limits a further exploration towards higher values of $\delta$.

\subsection{Spanwise particle motion}\label{results_spanwise}
\begin{figure*}
    \includegraphics[width=\textwidth]{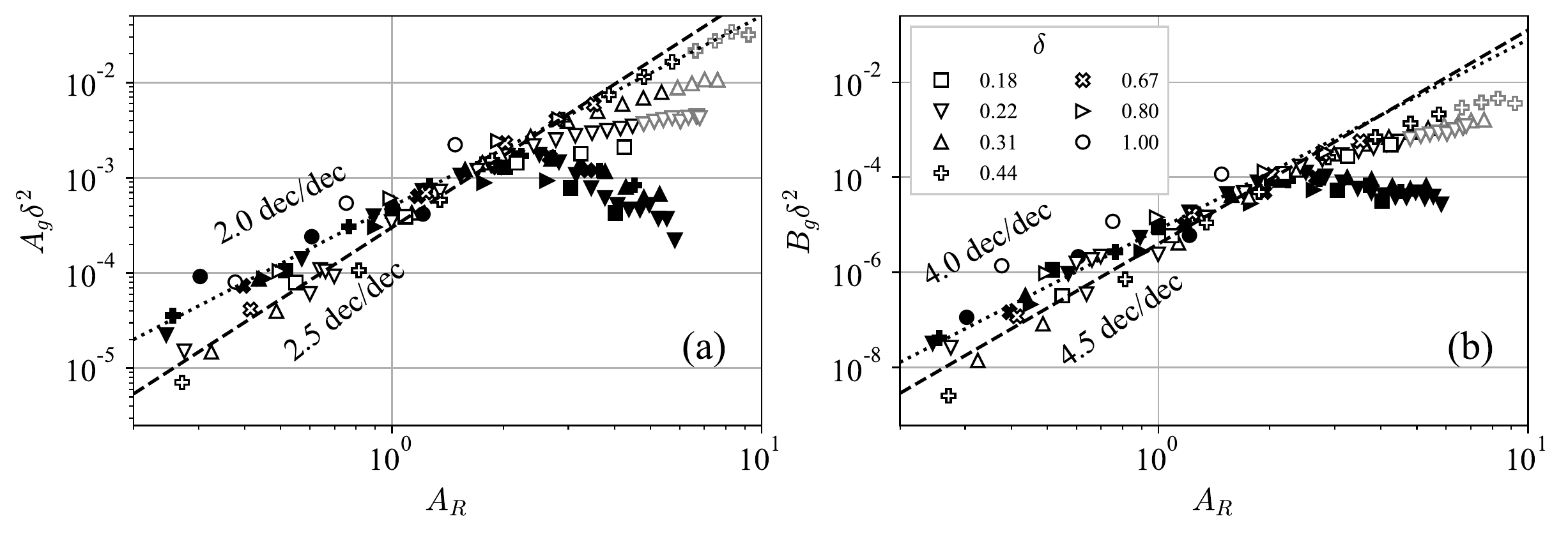}
    \caption{The amplitude of (a) the oscillation of the gap at twice the driving frequency $A_g$ and (b) the oscillation of the gap at four times the driving frequency $B_g$ as a function of the relative particle excursion length $A_R$, for $s=7.5$. The symbols are identical to those in Fig.~\ref{fig:figure5}. The amplitudes have been multiplied by $\delta^2$ such that the data for $A_R\lesssim2$ collapses. Two distinct regimes, with a transition around $A_R\approx2$, are found for both systems. The dotted and dashed lines correspond to Eqs.~\eqref{eq:Ag_Bg_scaling_osc}~and~\eqref{eq:Ag_Bg_scaling}, respectively, with their slopes annotated in the figure.}
    \label{fig:figure8}
\end{figure*}

In both systems, the particles oscillate relative to each other, perpendicularly to the bulk flow. This oscillation of the gap can be characterized in terms of the (normalized) amplitudes of the oscillations that occur, primarily, at twice and four times the driving frequency. These amplitudes are denoted by $A_g$ and $B_g$, respectively, as defined in Appendix~\ref{fittingequations}. When $A_R\lesssim2$, these amplitudes typically decrease with $\delta$ and increase with $A_R$. Furthermore, when scaled with $\delta^2$, as shown in Fig.~\ref{fig:figure8}, the amplitudes partially collapse onto a line when plotted as a function of $A_R$. 

Specifically, for $A_R\lesssim2$, the data for the oscillating box (for which $A_R=A_r$) is described by 
\begin{subequations}\label{eq:Ag_Bg_scaling_osc}
\begin{align}
    A_g &= C_A\left(\frac{A_R}{\delta}\right)^{2}, \\
    B_g &= C_B \left(\frac{A_R}{\delta}\right)^{2}A_R^{2},
\end{align}
\end{subequations}
with $C_A\approx 5\times10^{-4}$ and $C_B\approx 8\times10^{-6}$.
The data for the oscillating channel flow is described by similar relations
\begin{subequations}\label{eq:Ag_Bg_scaling}
\begin{align}
    A_g &= C'_A \left(\frac{A_R}{\delta}\right)^{2}A_R^{0.5}, \\
    B_g &= C'_B \left(\frac{A_R}{\delta}\right)^{2}A_R^{2.5},
\end{align}
\end{subequations}
with $C'_A\approx 3\times10^{-4}$ and $C'_B\approx 4\times10^{-6}$. Both scalings in Eq.~\eqref{eq:Ag_Bg_scaling} contain an additional factor $A_R^{0.5}$ compared to the scalings for the oscillating box. Equations \eqref{eq:Ag_Bg_scaling_osc} were presented previously by \citet{VanOverveld2022}, but the scaling for $B_g$ had an additional factor $\delta^{-0.5}$. Based on the available data, both variations are plausible. Here, we have chosen for the version in Eq.~\eqref{eq:Ag_Bg_scaling_osc}, since it yields an identical difference in functional dependency for $A_g$ and $B_g$ between the two systems.

For $A_R\gtrsim2$, the data for the oscillating box decreases in a scattered manner due to the widening of the gap as $L\approx 3.0\delta^{1.5}+0.03A_R^3$ ~\citep{VanOverveld2022}. As the distance between the particles increases, their instantaneous interactions become weaker, leading to smaller amplitudes. Analogously, the increase of $A_g$ and $B_g$ for the oscillating channel flow is due to the particles staying in each other's vicinity. In this system, the gap does not widen as drastically, as seen in Fig.~\ref{fig:figure5}. The increase in the instantaneous particle-fluid interactions with $A_R$ then results in larger oscillations of the gap.

\section{\label{sec:resultsDensityRatio}Effects of variations in the density ratio}
\subsection{Effect on particle dynamics}\label{results_densitydynamics}
The previous work for the oscillating box by \citet{VanOverveld2022} indicated that the mean state of the system (including the mean gap value $L$) is governed only by $\delta$ and $A_R$. Variation of the density ratio $s$ affected only the value of $A_R$ and hence lead to the same scalings and proportionality constants. 
Contrarily, we have hypothesized that the mean state of the oscillating channel flow has an additional degree of freedom (see sections~\ref{sec:introduction}~and~\ref{sec:formulation_particle}). 
Here, we show how this extra degree of freedom affects the equilibrium state by varying the particle-fluid density ratio $s$. We present the results from simulations with $s$ equal to $2.65$, $4.00$, $6.00$, and $7.50$. The value $s=2.65$ is commonly used for sediment transport\citep{Mazzuoli2016}, while $s=7.50$ corresponds to stainless-steel spheres in water-like fluids \citep{Klotsa2007}. In all other aspects, the simulations are identical to those from section~\ref{results_mean} with $\delta\approx0.22$. This particular value is chosen because the simulations in this part of the parameter space have a relatively low computational cost, allowing for an extensive scan over values of $s$ and $A_R$.

\begin{figure}
    \includegraphics[width=0.48\textwidth]{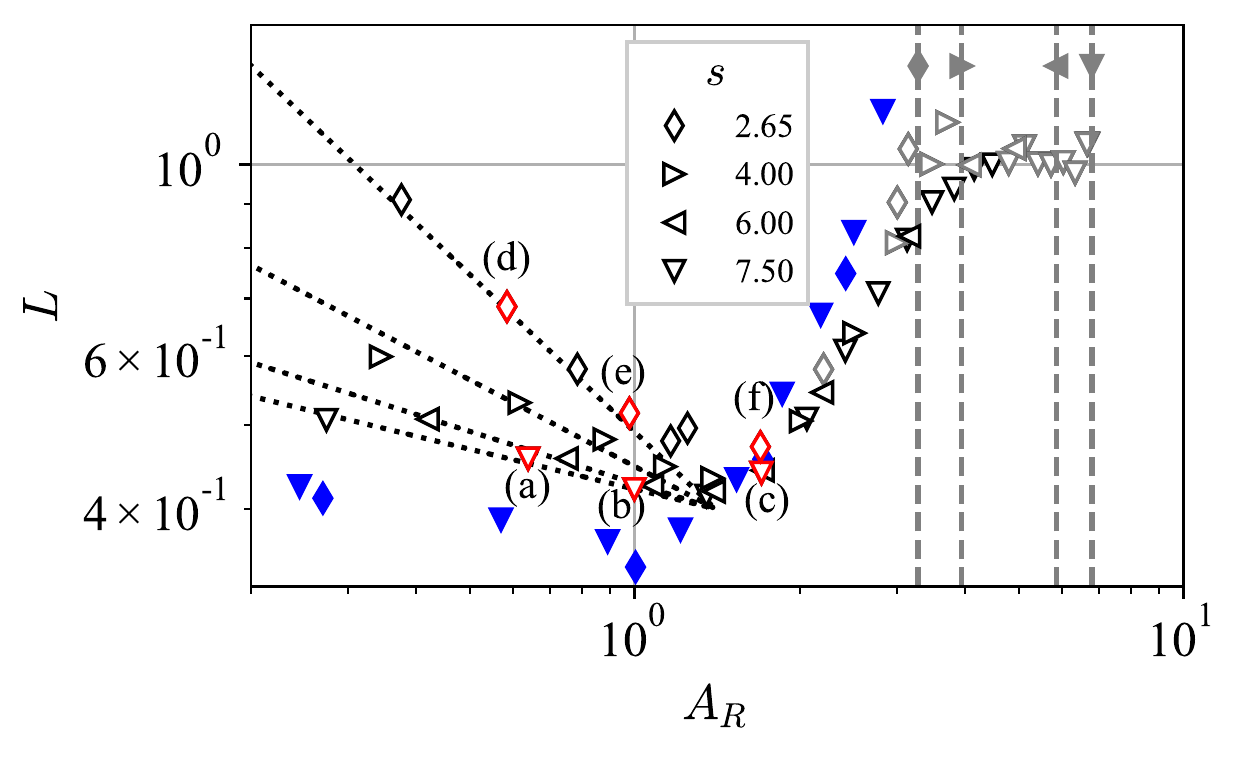}
    \caption{The mean value of the gap $L$ as a function of $A_R$, for $\delta\approx0.22$ and different values of $s$, for the oscillating channel flow (black and red). The gray symbols indicate the simulations in which particles lose contact with the bottom, while the dashed lines indicate above which values of $A_R$ the pair becomes unstable and the particles drift apart. The dotted lines correspond to Eq.~\eqref{eq:L_s_Ar} for each value of $s$. The characters (a-f) are placed near the (red) symbols that correspond to subfigures in Fig.~\ref{fig:figure10}, where flow fields are shown. The blue symbols correspond to the oscillating box data for $\delta\approx0.22$ and $s=2.65$ (diamonds) or $s=7.5$ (downward pointing triangles).}
    \label{fig:figure9}
\end{figure}

The mean values of the gap $L$ as a function of $A_R$ are shown in Fig.~\ref{fig:figure9}. For $A_R\lesssim1$, the mean gap decreases with increasing relative excursion length. The same effect is previously also seen for a range of $\delta$-values ($\delta\approx0.22$, $0.31$, and $0.44$) in Fig.~\ref{fig:figure5}. In Fig.~\ref{fig:figure9}, the gradient of the slope becomes more negative when $s$ is small, i.e. for lighter particles. Overall, the mean gap approximately follows  
\begin{equation}\label{eq:L_s_Ar}
    L \approx \left(0.4\right)^{s/(s-1)}\left(\frac{3.5}{ A_R}\right)^{1/(s-1)}
\end{equation}
where the numbers are empirically determined. The set of dotted lines in Fig.~\ref{fig:figure9} shows that this relation indeed describes the data. Note that the symbols and lines converge at $A_R\approx1.4$. The explicit dependence of $L$ on both $A_R$ and $s$ in Eq.~\eqref{eq:L_s_Ar} is a significant difference with the oscillating box\citep{VanOverveld2022}. In that system, the values of $L$ vary by less than $0.1$ for $A_R\lesssim1$, as shown by the blue symbols in Fig.~\ref{fig:figure9}. In other words, the mean gap is effectively only a function of $\delta$ in the viscous-dominated regime ($A_R\lesssim1$).

For $1.4\lesssim A_R\lesssim3.0$, the data from the oscillating channel flow in Fig.~\ref{fig:figure9} collapse for all density ratios without any rescaling. In this range, the value of $L$ rapidly increases with $A_R$ up to $L\approx1.0$ when $A_R\approx3.0$. At the lower end of the collapse, around $A_R\approx1.4$, the mean gap has a minimum at $L\approx0.41-0.46$ for each value of $s$. Note that the typical variations in $L$ (between $0.4$ and $1.0$) are relatively small compared to those found for the oscillating box in Fig.~\ref{fig:figure5} (between $0.5$ and $3.0$).

For $A_R\gtrsim3.0$ in Fig.~\ref{fig:figure9}, the data diverge. For low values of $s$ ($s=2.65$), the mean gap increases most with $A_R$, whereas for higher values ($s=7.50$), the mean gap remains at approximately $1$. Note that in this part of the parameter space, the particles lose contact with the bottom during as much as 25-50\% of each oscillation period. Upon increasing $A_R$, the lightest particles ($s=2.65$) get affected first, because the ratio between the upward lift force and the net downward gravitational force ($\sim(A_R/A)\Gamma\sqrt{\delta A}/(s-1)$; see Sec.~\ref{results_streamwise}) is higher for lower $s$ values. 
When the particles get lifted from the bottom, the particle pair can become unstable, after which the particles drift apart in both the streamwise and spanwise directions. The gray dashed lines indicate the minimum values of $A_R$ at which the pairs become unstable for each value of $s$. Even in elongated domains ($15\times40$) and starting close to the expected equilibrium configuration, the particles drift apart over typically $10-50$ oscillations.

\begin{figure*}
    \includegraphics[width=\textwidth]{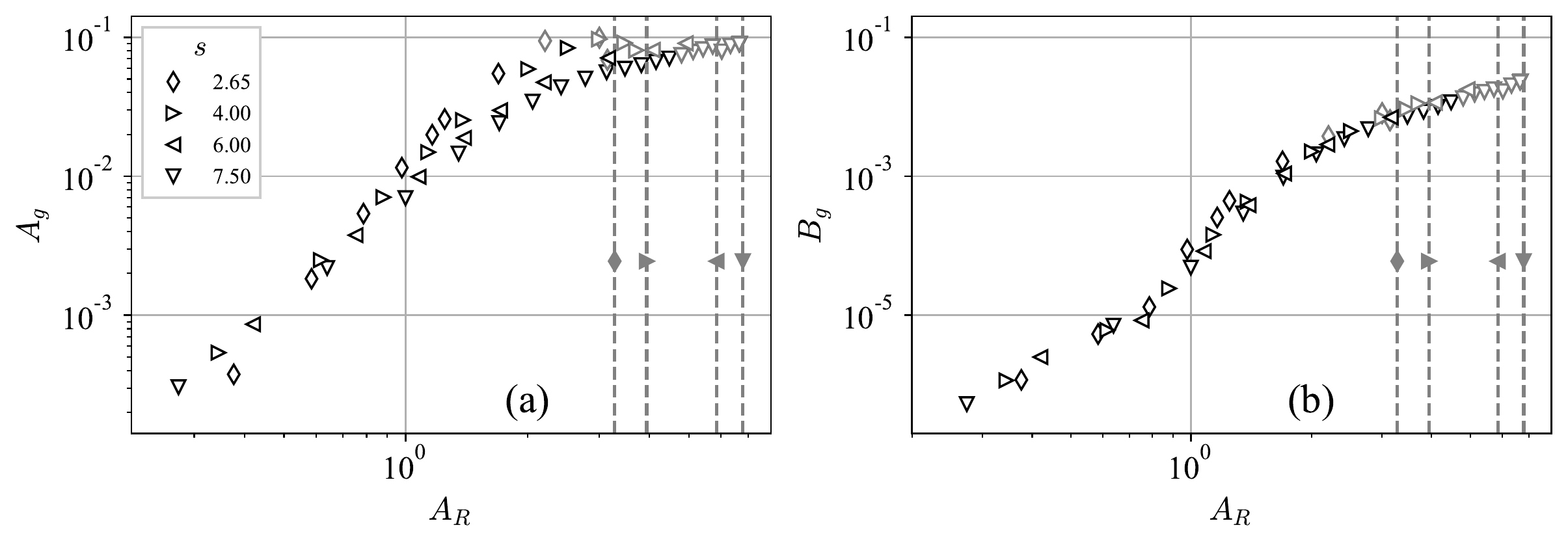}
    \caption{The amplitudes of the oscillation of the gap (a) $A_g$ and (b) $B_g$ as a function of $A_R$, for $\delta\approx0.22$ and different values of $s$. The symbols are identical to those in Fig.~\ref{fig:figure9}. The dashed lines indicate the value of $A_R$ at which the pair becomes unstable and the particles drift apart.}
    \label{fig:figure10}
\end{figure*}

In addition to the mean gap, we consider the oscillation amplitudes of the gap at twice and four times the driving frequency, $A_g$ and $B_g$, respectively, in Fig.~\ref{fig:figure10}. 
For $1\lesssim A_R\lesssim3$, the values of both $A_g$ and $B_g$ increase by approximately a factor $2$ when $s$ decreases from $7.50$ to $2.65$, at otherwise equal value of $A_R$. So, for these $A_R$ values, $s$ only affects the oscillation of the gap and not the mean gap itself. Contrarily, for $A_R\lesssim1$, the data of both $A_g$ and $B_g$ collapse onto a single curve quite well, without any rescaling. This means that, in this regime, the oscillation of the particles in the spanwise direction is not sensitive to variations in $s$. 

According to Figs.~\ref{fig:figure9}~and~\ref{fig:figure10}, there are simulations (e.g. (a) and (d) in Fig.~\ref{fig:figure9}) for which the streamwise and spanwise particle translations (in terms of $A_R$, $A_g$ and $B_g$) are similar, but the equilibrium configuration (in terms of $L$) is not. This means that the change in $L$ is due to a different physical mechanism. There are two mechanisms that can play a role: the shear between bottom and fluid (addressed in section~\ref{results_densityflowfields}) and the rotation of the particles (addressed in this section).

\begin{figure}
    \includegraphics[width=0.48\textwidth]{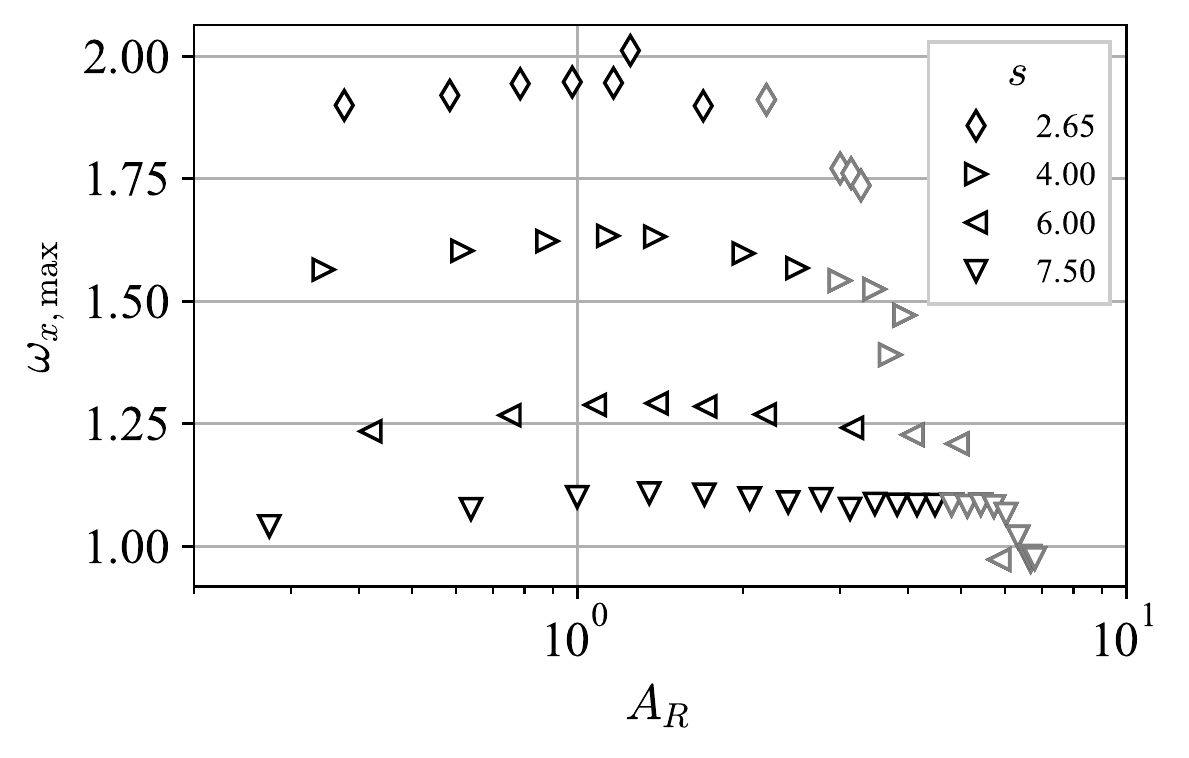}
    \caption{The maximum (dimensionless) angular rotation of the particles around the $x$-axis $\omega_{x,\mathrm{max}}$ as a function of $A_R$ for $\delta\approx0.22$ and different values of $s$. The symbols are identical to those in Fig.~\ref{fig:figure9}.}
    \label{fig:figure11}
\end{figure}

Even without particle-bottom friction, the particles can rotate around the $x$\nobreakdash-axis due to the vertical shear in the flow velocity. We quantify the rotation of the particles around the $x$\nobreakdash-axis using the maximum angular velocity $\omega_{x,\mathrm{max}}$. Note that this quantity is non-dimensionalized according to Eq.~\eqref{eq:nondim2}. The result is shown as a function of $A_R$ in Fig.~\ref{fig:figure11}. 

For most simulations, the value of $\omega_{x,\mathrm{max}}$ does not significantly vary with $A_R$, but does depend on $s$. Based on Fig.~\ref{fig:figure11}, the angular rotation scales approximately as $\omega_{x,\mathrm{max}}\sim s^{-0.5}$, whereas we expect it to scale as $\delta^2/s$, based on Eq.~\eqref{eq:particle_rot_dimless}. An explanation on the discrepancy between these scalings is absent. Still, for both the simulation data and the theory, we find that $\omega_{x,\mathrm{max}}$ is independent of $A_R$ (or $A$).

Increased particle rotation was previously correlated to an increase in the mean gap \citep{VanOverveld2022} and could thus explain the increase in $L$ at low $A_R$ values in Fig.~\ref{fig:figure9}. As verification, we performed additional simulations of the oscillating channel flow with $s=7.50$, $\Gamma=4.5$, and $\mu_c=[0.2,0.4]$. In these simulations, the particle-bottom friction delivers a torque that enhances the particle rotation up to $\omega_{x,\mathrm{max}}\approx 3.4$, which is significantly higher than the values in Fig.~\ref{fig:figure11}. The mean gap is approximately $0.1$ larger than when $\mu_c=0$, which indicates that particle rotation, induced by particle-bottom friction, increases the mean gap. More additional simulations in which the particle's moment of inertia is (artificially) multiplied or divided by a factor 3, yield values of $L$ that differ only $0.01$ from the base case.
The increase in $L$ due to particle rotation is thus only small compared to the increase due to lower values of $s$. Therefore, particle rotation is likely not the only physical mechanism that causes the increase in the mean gap.

\subsection{Effect on flow fields}\label{results_densityflowfields}
The differences in the equilibrium configuration, as shown in the previous section, are the result of differences in the particle-fluid interactions. Now that the particle dynamics are described, we investigate if the variation of the density ratio also leads to changes in the steady streaming flows. Note that changing the value of $s$, while keeping $A_R$ and $\delta$ constant, implies that $A$ changes accordingly, because there are only three degrees of freedom that define the system. For the flow, it is more relevant to consider $A$ instead of $s$, because $A$ is directly related to relative movement between bottom and (bulk) flow, and thus to the vertical shear in the velocity field.

\begin{figure*}
    \includegraphics[width=\textwidth]{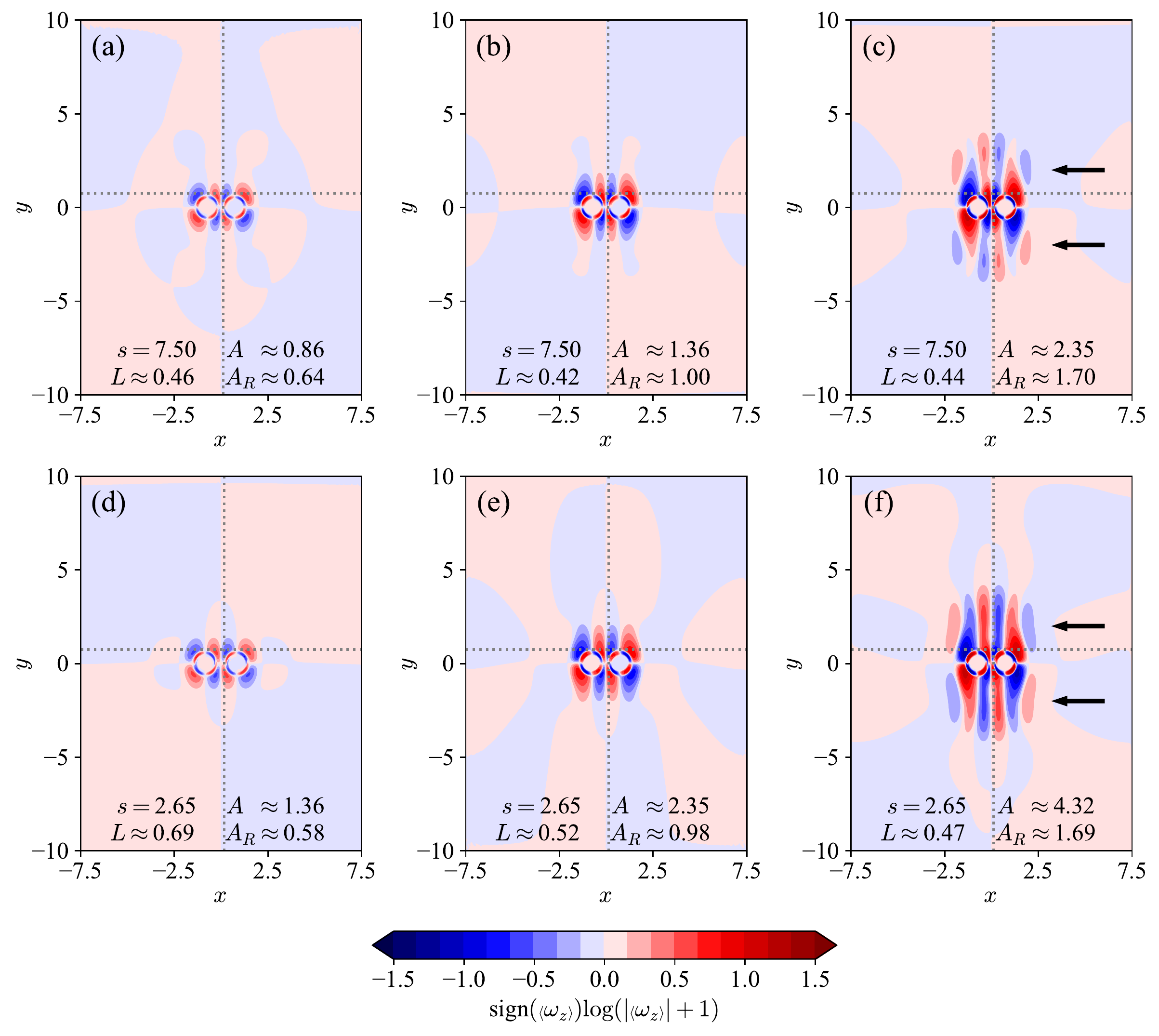}
    \caption{The time-averaged proxy for the vorticity in the $xy$-plane going through the centers of the particles for the oscillating channel flow with $\delta\approx0.22$, $s=7.50$ (top row) and $s=2.65$ (bottom row). The plots are aligned such that vertically adjacent plots have (approximately) the same value of $A_R$, while diagonally adjacent plots (b-d and c-e) have the same value of $A$. Specific parameter values are given within each subfigure. The black arrows in (c) and (f) point to the additional patches that emerge at high values of $A_R$. The vorticity along the dotted lines is plotted in Fig.~\ref{fig:figure13}.}
    \label{fig:figure12}
\end{figure*}

We visualize the time-averaged vorticity field in the horizontal plane at mid-particle height $z=1/2$ in Fig.~\ref{fig:figure12}. Horizontally aligned plots have the same value of $s$, while $A$ and $A_R$ both increase from left to right. Vertically aligned plots have similar values of $A_R$, whereas diagonally aligned plots have identical values of $A$.

The comparison in Fig.~\ref{fig:figure12} shows that the average vorticity strongly depends on $A_R$. To understand this, we recall that this vorticity is produced in the particle boundary layer, where the velocity shear scales with $A_R$~\citep{Riley1966}.
Upon increasing $A_R$, the patches close to the pair grow in magnitude and spatial extent. For the largest values of $A_R$ considered here (Figs.~\ref{fig:figure12}(c)~and~(f)), the patches are elongated in the oscillation direction and four additional patches emerge in the spanwise direction.
This confirms that the production of vorticity is determined by the velocity shear in the particle boundary layers, which scales with $A_R$~\citep{Riley1966}. 

The elongation in the $y$\nobreakdash-direction was previously observed in the oscillating box for $A_R\gtrsim1$, when the advection of vorticity becomes relatively important with respect to the dissipation \citep{VanOverveld2022}. However, the additional patches were not observed in the oscillating box system.

On the other hand, the average vorticity is hardly affected by variations in $A$ (or $s$) at constant value of $A_R$. For low $A_R$\nobreakdash-values (Figs.~\ref{fig:figure12}(a)~and~(d)), the patches close to the particles appear quite similar, even though $L$ varies by a factor $1.5$. For high $A_R$\nobreakdash-values (Figs.~\ref{fig:figure12}(c)~and~(f)), an increase of a factor $1.8$ in $A$ does affect the vorticity close to the particles. For large $A$, the patches are elongated further in the streamwise direction, and the vorticity magnitude inside them is higher, especially, in the parts further away from the pair. Note that despite these differences in the flow fields, the value of $L$ is not significantly different.

\begin{figure*}
    \includegraphics[width=\textwidth]{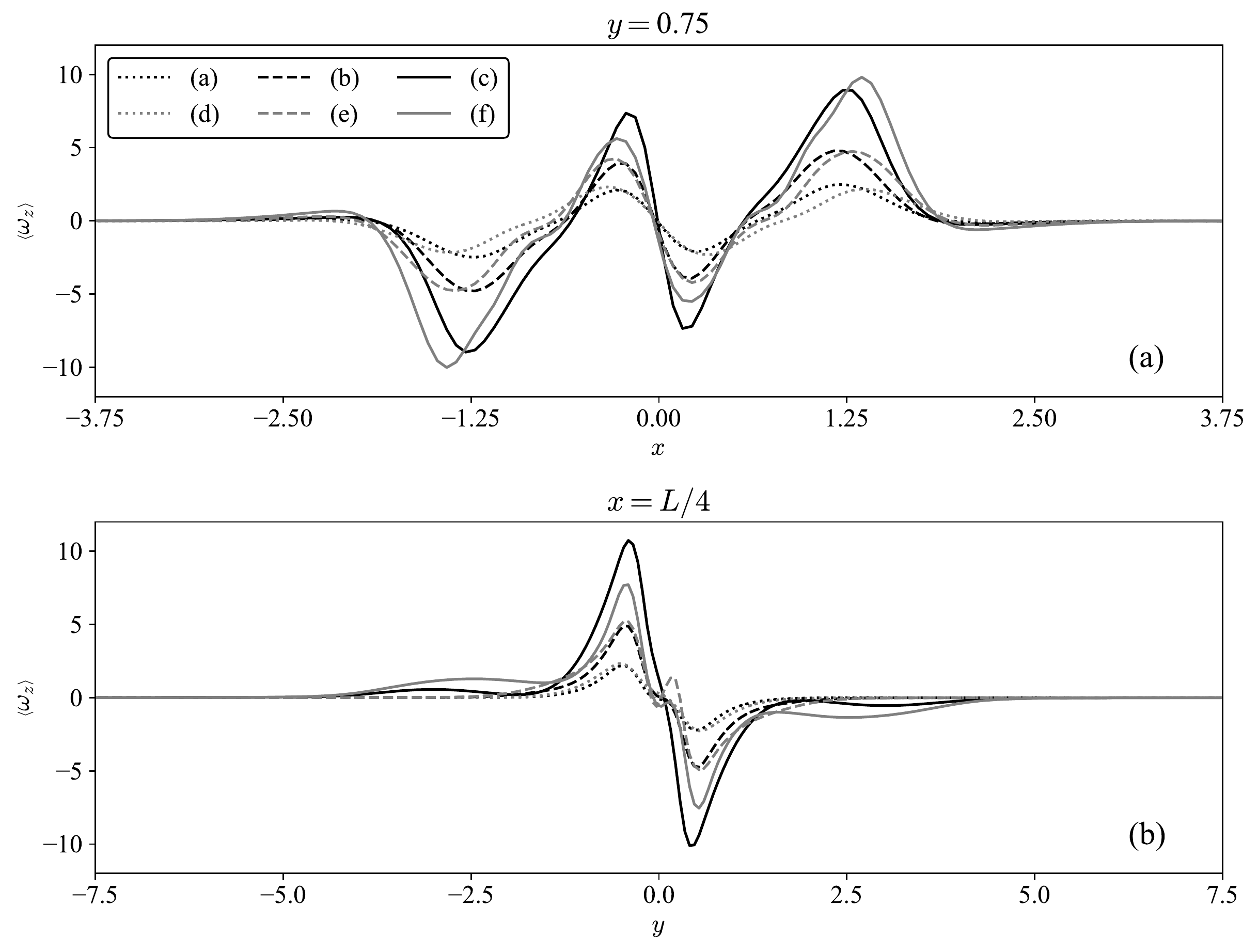}
    \caption{The average vorticity $\left<\omega_z\right>$ along the lines (a) $y=0.75$ and (b) $x=L/4$ for the six cases that are shown in Fig.~\ref{fig:figure12}. The black and gray curves correspond to the cases with $s=7.50$ and $s=2.65$, respectively. Note that the dashed curves have approximately the same magnitude and shape, especially close to the particles. The same holds for the dotted curves.}
    \label{fig:figure13}
\end{figure*}

To make a stronger quantitative comparison and better illustrate these points, we calculate the time-averaged vorticity $\left<\omega_z\right>$ (see Eq.~\eqref{eq:meanvorticity}) along the lines $y=0.75$ and $x=L/4$ (with values between $x\approx0.11$ and $0.17$). The results are shown in Fig.~\ref{fig:figure13}. The positions of the lines, as indicated in Fig.~\ref{fig:figure12}, include the vorticity patches that vary between the cases, while they exclude the symmetry axes of the configuration and the thin layer of vorticity around each particle. The lines at $x=L/4$ lie precisely between the symmetry axis ($x=0$) and the particle surface ($x=L/2$). Even though the exact position of the lines is arbitrary, the interpretation of the results is not sensitive to small changes in their positions. 

In Fig.~\ref{fig:figure13}, the two dotted curves (corresponding to Figs.~\ref{fig:figure12}(a)~and~(d)) have similar shapes and magnitudes. Especially close to the pair (around $x=0$ and $y=0$), the curves overlap. Further from the origin, the gray curve is shifted further outwards with respect to the black curve. The same observations hold for the dashed curves, which correspond to Figs.~\ref{fig:figure12}(b)~and~(e). So, for increasing values of $A$, at constant $A_R$, the time-averaged vorticity distribution is situated further away from the pair. The vorticity patches shown in Fig.~\ref{fig:figure12} are thus elongated in both the streamwise and spanwise direction.

\begin{figure}
    \includegraphics[width=0.485\textwidth]{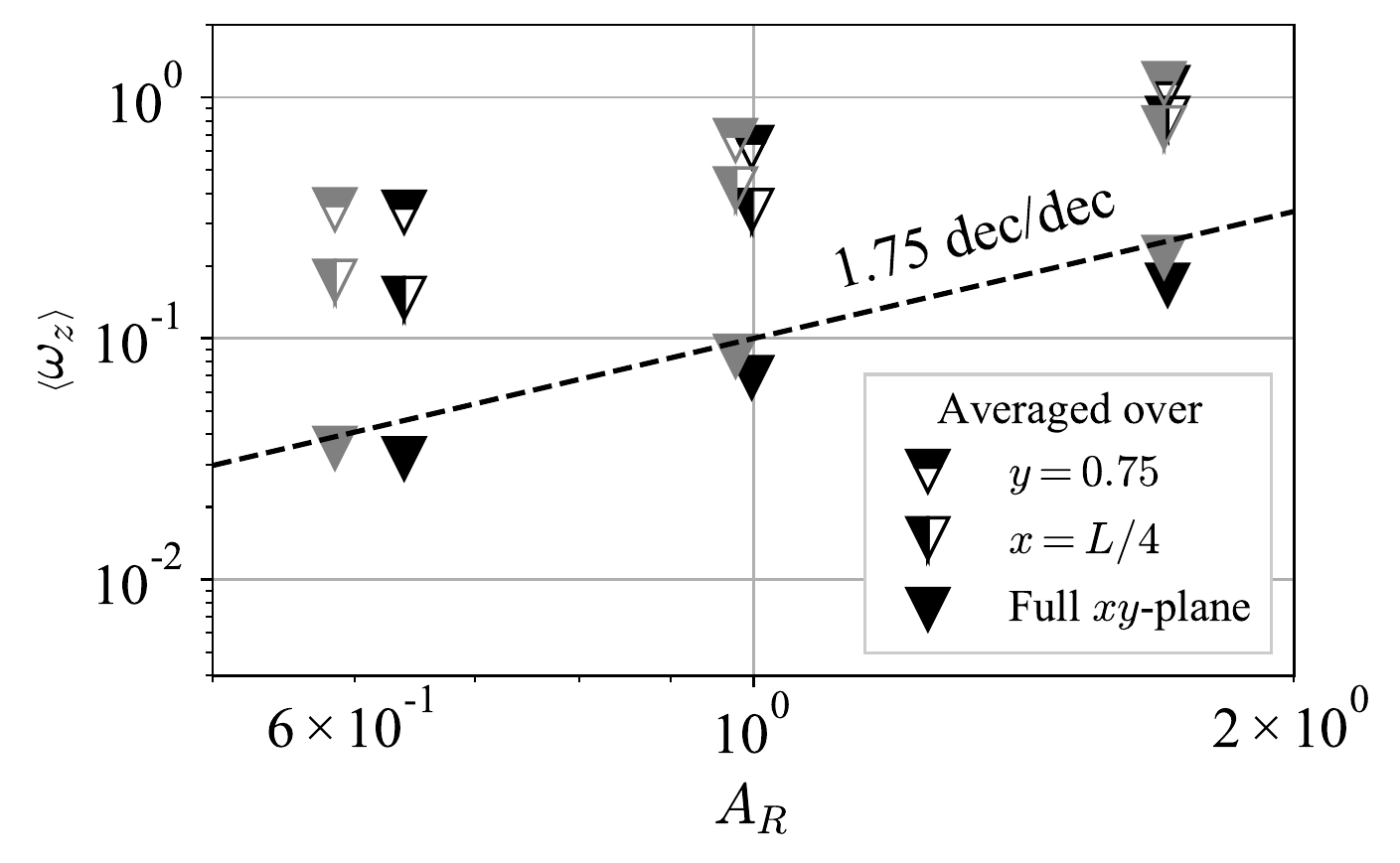}
    \caption{The magnitude of the time-averaged vorticity $\left<\omega_z\right>$ for the six cases shown in Fig.~\ref{fig:figure12}, averaged over the lines $y=0.75$ and $x=L/4$. Additionally, the average over the full $xy$-plane is shown, which is equivalent to $C$ as defined in Eq.~\ref{eq:avgmeanvorticity}. The black ($s=7.50$) and gray ($s=2.65$) symbols correspond to the black and gray curves in Fig.~\ref{fig:figure13}, respectively.}
    \label{fig:figure14}
\end{figure}

The solid curves in Fig.~\ref{fig:figure13} have, on top of the outward shift, also a significant difference in the vorticity magnitude. For example, the maximum vorticity of the two solid curves in Fig.~\ref{fig:figure13}(b) varies from a factor $0.5$ at $y\approx-2.5$, up to a factor $1.4$ at $y\approx-0.5$. These different magnitudes could be due to either redistribution or differences in the production of vorticity. Hence, we average the (absolute) vorticity along the two lines $y=0.75$ and $x=L/4$, and show the results as a function of $A_R$ in Fig.~\ref{fig:figure14}. In addition, we have included an average over the full horizontal plane, which is equal to $C$, as defined in Eq.~\eqref{eq:avgmeanvorticity}.

The (total) vorticity in the plane is strongly dependent on $A_R$, scaling approximately with $A_R^{1.75}$. This confirms that the production of vorticity is coupled to the velocity shear in the particle boundary layers, which scales with $A_R$~\citep{Riley1966}. 
Contrarily, variation of $A$ (or $s$) only has a small effect on the spatially averaged vorticity. For example, at $A_R\approx1.7$ (rightmost symbols in Fig.~\ref{fig:figure14}), the line-averaged vorticity varies by only 4\% and 9\%, while the value of $A$ nearly doubles. Variation of $A$, at constant $A_R$, does thus not affect the total amount of vorticity, but rather redistributes it over the horizontal plane. This redistribution can lead to differences in the particle-fluid interactions and subsequently into differences in the mean gap. More specifically, we have seen in Fig.~\ref{fig:figure12} that for larger values of $A$, the vorticity is spread further away from the particles and the mean gap increases.

So, for the oscillating channel flow, the two excursion lengths $A_R$ and $A$ can be assigned to different physical mechanisms. While $A_R$ is related to both the production and advection of vorticity, $A$ sets the vertical shear and can further enhance the transport while keeping the total amount of vorticity unaffected. For the oscillating box, this latter mechanism is absent \citep{VanOverveld2022}.

\section{\label{sec:discussion}Discussion}
In section~\ref{sec:resultsDensityRatio}, we have shown that the mean state of the system (e.g. $L$) is affected by particle rotation. The rotation is due to a net torque on the particle, as a result of velocity gradients or particle-bottom friction, represented by the first and second term on the right-hand side of Eq.~\eqref{eq:particle_rot_dimless}, respectively. However, for most of our simulations, we have neglected particle-bottom friction; an assumption that we discuss first. Then we extend the discussion to the influence of the density ratio on the mean gap and the relevance of our results for different parts of the parameter space.

Particle-bottom friction is characterized by the dimensionless parameter $\mu_c/\Gamma$ (see Eqs.~\eqref{eq:particle_trans_dimless}~and~\eqref{eq:particle_rot_dimless}), given that the hydrodynamic lift can be neglected. \citet{VanOverveld2022} showed that 
simulations where the particle-bottom friction is absent ($\mu_c=0$) agree well with experiments at relatively high frequencies\citep{Klotsa2007}, such that the oscillatory acceleration is larger than the gravitational acceleration ($\Gamma>1$). In such experiments, strong particle chains have been found \citep{Klotsa2009}. Our simulation results, with $\mu_c=0$, thus also correspond to this particle-chain regime.
However, the friction needs to be incorporated to accurately simulate systems where $\Gamma$ is smaller, e.g. for the rolling-grain ripples \citep{Mazzuoli2016}. 

Alternatively, particle rotation can also be caused by velocity gradients in the Stokes boundary layer, as shown in section~\ref{results_densitydynamics}. We can use these results to better understand situations where particle-bottom friction cannot be neglected. Most likely, the cause of the rotation is not important, but the subsequent particle-fluid interactions are. Note that for the oscillating box, there is no Stokes boundary layer that can induce a net torque on the particle. The moment of inertia is then not a relevant quantity for the mean equilibrium state of the system \citep{VanOverveld2022}.

Additionally, we have shown in section~\ref{results_densityflowfields} through variation in $s$, at constant values of $A_R$ and $\delta$, that $A$ is an important quantity for the distribution of the period-averaged vorticity over the horizontal plane. We have shown that the mean gap is larger when the vorticity is smoothed out over a larger part of the domain. This indicates that the (average) attraction between the particles is larger when the vorticity is concentrated closely around the particles. This interpretation also agrees with results from the oscillating box, where for small values of $\delta$, the vorticity is concentrated in a thin layer around the particles\citep{VanOverveld2022}. The mean gap in these cases is also small and scales as $L\sim\delta^{1.5}$.

Besides the aforementioned differences, the other parameter values in environmental settings can be different than in our simulations. Hence, we discuss the accuracy and applicability of our results in the different parts of the parameter space.
In particular, when both $A_R$ and $\delta$ are small, $L$ converges to the values found for the viscous regime in the oscillating box. The Stokes boundary layer does not play a large role in this part of the parameter space and we indeed expect the two systems to become equivalent.

Only when $\delta$ is sufficiently large, i.e. $\delta^{0.7}\gtrsim A_R^{1.9}$ (see Eq.~\ref{eq:C_AR_delta}), does the viscous dissipation become important relative to the production of vorticity, such that the formation of steady streaming flow is suppressed. In such a case, the residual flow is almost zero and the subsequent interaction with the particles is weak. For the cases where the steady streaming flow is extremely weak, $L$ changes by only $\sim10^{-3}$ per oscillation period and typically thousands of periods are needed until a quasi-steady state is reached, even when starting close to the equilibrium configuration. Therefore, we expect that in this part of the parameter space, particle pairs are extremely rare in environmental or laboratory settings. Due to their weak interaction, the pair stability would be too sensitive to small perturbations due to e.g. bottom roughness.

On the other hand, we have consistently found steady streaming flows for $A_R>1$ in the oscillating channel flow, at every value of $\delta$. Specifically for small $\delta$, the particles stay close together, where they would drift apart in the oscillating box. For larger $\delta$, as is more common in environmental settings, the particle-particle interactions are sufficiently large to reach an equilibrium state with a typical time scale of $\sim100$ oscillation periods, based on values of the fitting parameter $\tau$ (discussed in Appendix~\ref{fittingequations} in Eq.~\eqref{gap_exp}).
For the formation of patterns, however, different time scales could be relevant. For example, the alignment of the pairs from parallel to perpendicular to the oscillation direction could occur in fewer oscillation periods. Additionally, the presence of other particles in denser systems could accelerate the ordering processes, illustrated by e.g. the formation of particle chains in approximately $20$ oscillation periods by \citet{Mazzuoli2016}. 

In short, the ordering mechanisms for the oscillating box\citep{Klotsa2007,Klotsa2009,Wunenburger2002} and for the initiation of rolling-grain ripples\citep{Mazzuoli2016} seem qualitatively similar, albeit with a large variation in strength and dependence on the parameters governing the problem. The typically weak ordering mechanism at large $\delta$-values (at constant $A_R$) would allow for, for example, more tortuosity in longer sediment chains and the presence of defects in the pattern. Both these effects are commonly observed in simulations and experiments in this part of the parameter space \citep{Mazzuoli2016,Hwang2008,Shibata1993}. 

\section{\label{sec:conclusions}Conclusions}
In this paper, we have described the dynamics of a particle pair in an oscillating channel flow and compared them to those of a particle pair in an oscillating box. The results are obtained using direct numerical simulations, where the Navier-Stokes equations are solved in a double-periodic domain with a no-slip bottom. The motion of the particles and their interaction with the fluid phase is accounted for using the immersed boundary method.

The equilibrium states of the two systems have marked differences in the dynamics of the particles and the steady-streaming flow around them. In absence of particle-bottom friction, the oscillating box is governed by two dimensionless parameters: the normalized relative excursion length $A_R$ and the normalized Stokes boundary layer thickness $\delta$. The particle dynamics in the oscillating channel flow are governed by an additional dimensionless parameter. The extra degree of freedom is represented by either the particle-fluid density ratio $s$, which controls the amount of particle rotation, or the excursion length of the bulk flow $A$, which controls the distribution of the period-averaged vorticity over the horizontal plane. This latter mechanism has a major influence on the mean gap between the particles.

In general, the oscillating box and oscillating channel flow are different systems, even though the ordering mechanisms in both systems are qualitatively similar. Results obtained in one system cannot be directly translated to the other. Only in a limited part of the parameter space, when both $\delta$ and $A_R$ are small, can the findings for the oscillating box be applied to the oscillating channel flow.
It is thus of paramount importance to discriminate between the two systems and regions of the parameter space when comparing seemingly similar phenomena such as the particle chains described by \citet{Klotsa2009} and the rolling-grain ripples studied by \citet{Mazzuoli2016}.

\section*{Supplementary Material}
See supplementary material for the animations illustrating the three-dimensionality of the structures shown in Fig.~\ref{fig:figure2}.

\begin{acknowledgments}
We like to thank the staff in charge of the Reynolds cluster at the Delft University of Technology.
\end{acknowledgments}

\section*{Data Availability Statement}
The data that support the findings of this study are openly available in 4TU.ResearchData at http://doi.org/10.4121/20375364.

\appendix
\section{Analytical velocity profiles}\label{analyticalvelocityprofiles}
We consider an unbounded oscillating flow over a horizontal, fixed bottom at $z=0$. A harmonically oscillating pressure gradient $\nabla p_e=-\cos(2\pi t)\hat{\boldsymbol{y}}$ drives the flow with excursion length $A$ and angular frequency $2\pi$ (both dimensionless). This problem is a variation on Stokes second problem, given by 
\begin{equation}\label{stokesproblem}
    \frac{1}{2\pi}\frac{\partial \boldsymbol{u}}{\partial t} = \frac{1}{2}\delta^2\nabla^2\boldsymbol{u} + \cos(2\pi t)\hat{\boldsymbol{y}},
\end{equation}
with solution
\begin{equation}\label{velocity_analytical_unbounded}
    \boldsymbol{u} = \left[\sin(2\pi t)-e^{-z/\delta}\sin{\left(2\pi t - \frac{z}{\delta}\right)}\right]\hat{\boldsymbol{y}}.
\end{equation}
The fluid excursion at the height of the particle center is
\begin{eqnarray}\label{amplitude_analytical_unbounded_particleheight}
    \left.\boldsymbol{x}_f\right|_{z=1/2} &=& 2\pi A \int\left.\boldsymbol{u}\right|_{z=1/2}dt\\
    &=& -A\left[\cos(2\pi t)-e^{-1/2\delta}\cos{\left(2\pi t - \frac{1}{2\delta}\right)}\right]\hat{\boldsymbol{y}},\nonumber
\end{eqnarray}
where the factor $2\pi A$ comes from the different scales used in the nondimensionalization of $\boldsymbol{u}$ (with $A'\omega$) and $x$ (with $D$), see Eq.~\eqref{eq:nondim}.
The vertical shear is
\begin{equation}\label{velocity_shear}
    \frac{\partial \left(\boldsymbol{u\cdot}\hat{\boldsymbol{y}}\right)}{\partial z} = -\frac{1}{\delta}e^{-z/\delta}\left[\sin{\left(2\pi t - \frac{z}{\delta}\right)}+\cos{\left(2\pi t - \frac{z}{\delta}\right)}\right],
\end{equation}
such that, due to the non-dimensionalization of the velocity (see Eq.~\ref{eq:nondim}), the typical (dimensionfull) shear rate is given by $\dot\gamma=A\omega/\delta$. 

For an oscillating flow in a channel bounded by horizontal plates at $z=0$ and $z=H$, the solution to Eq.~\ref{stokesproblem} is more complex. It is given by
\begin{equation}\label{velocity_analytical_bounded}
    \boldsymbol{u} = \Re\left[\sin{(2\pi t)} +  \frac{\cosh{\left[(1+i)(2z-H)/2\delta\right]}}{\cosh{\left[(1+i)H/2\delta\right]}}ie^{2\pi it}\right]\hat{\boldsymbol{y}},
\end{equation}
in which $\Re\left[\dots\right]$ denotes the real part of the expression between the brackets. The ratio $H/\delta$ is related to the Womersley number, which is used in e.g. the description of pulsatile blood flow \citep{womersley1955}. In the limit of $H/\delta\gg1$, the velocity in the range $0\leq z\leq H/2$ converges to that of equation Eq.~\eqref{velocity_analytical_unbounded}.

\section{Analytical particle trajectory}\label{BBOmodel}
Here, we present a derivation of the analytical particle trajectories, given in dimensionfull form, such that it matches the commonly used formulation, e.g. by \citet{corrsin1956}.
We consider an external pressure gradient
\begin{equation}\label{eq:pressure}
    \boldsymbol{\nabla}' p' = -A'\omega^2\rho_f e^{i\omega t'}\boldsymbol{\hat{y}},
\end{equation}
that drives an oscillating flow, such that the velocity field is described by
\begin{equation}\label{eq:undisturbedflow}
    \boldsymbol{u}'_f = -A'\omega ie^{i\omega t'}\boldsymbol{\hat{y}}.
\end{equation}
We assume that the motion of a spherical particle immersed in such a flow is described by 
\begin{equation}\label{eq:particlemotion}
    \boldsymbol{u}'_s = -A'_s\omega ie^{i(\omega t'+\phi)}\boldsymbol{\hat{y}},
\end{equation}
with $A_s$ and $\phi$ the excursion length and phase lag of the particle, respectively. If the particle is sufficiently small, the flow can be approximated by the undisturbed flow field of Eq.~\ref{eq:undisturbedflow}. Only cases at low Reynolds numbers are considered, such that Stokes drag applies. The translation of the particle is then described by the Basset-Boussinesq-Oseen (BBO) equation, based on fundamental work from each of the authors \citep{basset1888,boussinesq1903,oseen1927}. We use the form similar to that given by \citet{corrsin1956}:
\begin{eqnarray}\label{eq:BBO}
    \frac{\pi}{6}\rho_sD^3\frac{d\boldsymbol{u}'_s}{dt'} = &&3\pi\rho_f f\nu D(\boldsymbol{u}'_f-\boldsymbol{u}'_s) - \frac{\pi}{6}D^3\boldsymbol{\nabla}' p' \nonumber\\
    &&+ \frac{\pi}{12}\rho_fD^3\frac{d(\boldsymbol{u}'_f-\boldsymbol{u}'_s)}{dt'} \\
    &&+ \frac{3}{2}D^2\rho_f\sqrt{\pi f\nu}\int_{t_0}^{t'}\frac{1}{\sqrt{t'-\tau}}\frac{d(\boldsymbol{u}'_f-\boldsymbol{u}'_s)}{dt'}d\tau. \nonumber
\end{eqnarray}
This equation equals the force on the particle to the sum of the Stokes drag, the pressure gradient in the undisturbed flow, the added mass, and the Basset history force. An a priori unknown factor $f$ is added to the terms that model the viscous drag, i.e. the Stokes drag and the Basset history force. This factor accounts for the enhanced drag on a sphere in vicinity of a wall\citep{goldman1967}. We expect $f=1$ for an unbounded domain and $f>1$ for the systems considered in this study. The added mass likely also changes due to the presence of a wall, but this effect will not be taken into account here.

We now use expressions Eqs.~\eqref{eq:pressure}, \eqref{eq:undisturbedflow}, and \eqref{eq:particlemotion} as ansatz in Eq.~\eqref{eq:BBO}. For the treatment of the Basset force, we assume that the system has reached a quasi-steady state, such that we can use the limit 
\begin{eqnarray}
    \lim_{t_0\rightarrow -\infty}\int_{t_0}^{t'}\frac{e^{i\omega \tau}d\tau}{\sqrt{t'-\tau}} &&= \lim\limits_{t_0\rightarrow -\infty} \sqrt{\frac{\pi}{i\omega}}\mathrm{erf}\left(\sqrt{i\omega\left(t'-t_0\right)}\right)e^{i\omega t'} \nonumber\\
    &&= \left(1-i\right)\sqrt{\frac{\pi}{2\omega}}e^{i\omega t'},
\end{eqnarray}
in which $\mathrm{erf}(\dots)$ is the error function.
By equating the real and imaginary parts of Eq.~\eqref{eq:BBO}, expressions for the particle excursion length 
\begin{equation}
    \frac{A_s}{A} = \sqrt{\frac{(9f\delta)^2\left(2f\delta+1\right)^2+9\left(3f\delta+1\right)^2}{(9f\delta)^2\left(2f\delta+1\right)^2+\left(9f\delta+2s+1\right)^2}}
\end{equation}
and phase lag
\begin{equation}
    \tan\left(\phi\right) = \frac{-18f\delta\left(2f\delta+1\right)\left(s-1\right)}{(9f\delta)^2\left(2f\delta+1\right)^2+\left(9f\delta+2s+1\right)\left(9f\delta+3\right)}
\end{equation}
are found. Finally, using Eq.~\eqref{eq:amplitudes}, we obtain an expression for the excursion length of the particle relative to the ambient flow 
\begin{equation}\label{eq:Fsfd}
    F(s,f,\delta) \equiv \frac{A_R}{A} = \frac{2(s - 1)}{\sqrt{(9f\delta)^2(2f\delta+1)^2+(9f\delta+2s+1)^2}}.
\end{equation}

\section{Equations used for least squares fitting}\label{fittingequations}
The particle amplitude in the lab frame $A_s$ is obtained by fitting the function 
\begin{equation}
    f_{A_s}(t) = y_0 + v_0t + A_s\sin{(2\pi t+\theta)},
\end{equation}
to the particle position in the streamwise direction. The fitting parameters $y_0$ and $v_0$ correct for transient effects in the mean position.

Likewise, the mean gap $L$ is obtained from the distance between the particles, to which the function
\begin{eqnarray}\label{gap_exp}
    f_\mathrm{gap}(t) = && L + a e^{-t/\tau} + A\cos(4\pi t+\theta_1) \nonumber\\
    && + B\cos(8\pi t+\theta_2) + C\cos(2\pi t+\theta_3)
\end{eqnarray}
is fitted. The normalized amplitudes of the oscillation of the gap $A_g$ and $B_g$ are obtained similarly as $L$, using
\begin{eqnarray}
    f_\mathrm{osc}(t) = &&a+bt+ct^2 + A_g\cos(4\pi t+\theta_1) \nonumber\\
    &&+ B_g\cos(8\pi t+\theta_2) + C_g\cos(2\pi t+\theta_3).
\end{eqnarray}
The polynomial part is only used for interpolation over a few periods of the main oscillation.

\nocite{*}
\bibliography{references}

\end{document}